\documentclass[12pt]{article}
\usepackage{latexsym,amssymb}

\thicklines                         
\def\footnoteitem(#1)#2{
\begin{list}{#1}{\labelwidth4.0mm \leftmargin7.0mm
\labelsep2.5mm \rightmargin7.0mm \parsep0.5ex plus0.2ex minus0.1ex
\itemsep0ex plus0.2ex }
\item #2
\end{list}
}

\begin{document}

\long \def \blockcomment #1\endcomment{}

\def\arcosh{\mathop{\rm arcosh}\nolimits}
\def\leqx{\,\raisebox{-1.0ex}{$\stackrel{\textstyle <}{\sim}$}\,}

\newcommand{\be}{\begin{equation}}
\newcommand{\ee}{\end{equation}}
\newcommand{\ba}{\begin{eqnarray}}
\newcommand{\ea}{\end{eqnarray}}

\newcommand{\ie}{{\it i.e.}}
\newcommand{\cf}{{\it cf.}}
\newcommand{\etal}{{\it et al.}}
\newcommand{\etc}{{\it etc}}
\newcommand{\eg}{{\it e.g.}}

\newcommand\qt{{\widetilde q}}
\newcommand\ut{{\widetilde u}}
\newcommand\dt{{\widetilde d}}
\newcommand\qtd{{\qt^\dagger}}
\newcommand\qbar{{\overline q}}
\newcommand\Mbar{{\overline M}}
\newcommand\Mt{{\widetilde M}}
\newcommand\Mtb{{\overline\Mt}}
\newcommand\Abar{{\overline A}}
\newcommand\cL{{\cal L}}
\newcommand\Vt{{\widetilde V}}
\newcommand\tr{{\rm tr\,}}
\newcommand\sign{{\rm sign\,}}
\newcommand\mt{{\widetilde m}}
\newcommand\mtb{{\overline{\mt}}}
\newcommand\Det{{\rm det\,}}
\newcommand\alphabar{{\overline\alpha}}
\newcommand\betabar{{\overline\beta}}
\newcommand\Psibar{{\overline{\Psi}}}
\newcommand\cM{{\cal M}}
\newcommand\cMbar{{\overline{\cM}}}
\newcommand\cA{{\cal A}}
\newcommand\cAbar{{\overline{\cA}}}
\newcommand\phibar{{\hat{\phi}}}
\newcommand\chibar{{\overline{\chi}}}
\newcommand\str{{\rm str\,}}
\newcommand\sdet{{\rm sdet\,}}
\newcommand\cU{{\cal U}}
\newcommand\cV{{\cal V}}
\newcommand\cS{{\cal S}}
\newcommand\cG{{\cal G}}
\newcommand\cH{{\cal H}}
\newcommand\phit{{\widetilde\phi}}
\newcommand\diag{{\rm diag\,}}
\newcommand\Exp{{\rm exp}}
\newcommand\xbar{{\overline{x}}}
\newcommand\ybar{{\overline{y}}}
\newcommand\vx{{\vec x}}

\begin{titlepage}

\rightline{LA-UR-050107}
\rightline{UW/PT 05-02}
\vskip 5mm

\baselineskip=20pt plus 1pt
\vskip 0.5cm
\centerline{\LARGE Effective Theory for Quenched Lattice QCD
}
\vskip 0.5cm
\centerline{\LARGE and the Aoki Phase}
\vskip 1.0cm
\centerline{\large Maarten Golterman$^1$}
\centerline{\em  Department of Physics and Astronomy, 
San Francisco State University,}
\centerline{\em 1600 Holloway Ave, San Francisco, CA 94132, USA}
\vskip 0.5cm
\centerline{\large Stephen R. Sharpe$^2$}
\centerline{\em Department of Physics, University of Washington, Seattle,
WA 98195, USA}
\vskip 0.5cm
\centerline{\large Robert L. Singleton Jr.$^3$}
\centerline{\em Los Alamos National Lab., Group X-7, Los Alamos, NM 87545, USA}
\vskip 2.0cm
\baselineskip=12pt plus 1pt
\parindent 20pt
\centerline{\bf Abstract}
\textwidth=6.0truecm
\medskip

\frenchspacing

We discuss the symmetries of quenched QCD with Wilson fermions,
starting from its lagrangian formulation, 
taking into account the constraints needed for convergence of
the ghost-quark functional integral. We  
construct the corresponding chiral effective lagrangian,
including terms linear and quadratic in the lattice spacing.  
This allows us to study the phase
structure of the quenched theory, and compare it
to that in the unquenched theory. In particular we study
whether there may be an Aoki phase (with parity and flavor spontaneously
broken) or a first order transition line
(with no symmetry breaking but meson masses proportional to 
the lattice spacing), which are the two possibilities
in the unquenched theory.
The presence of such phase structure,
and the concomitant long-range correlations,
has important implications for numerical studies using both
quenched and dynamical overlap and domain-wall fermions.
We argue that the phase structure is qualitatively the same
as in the unquenched theory, with the choice between the
two possibilities depending on the
sign of a parameter in the low-energy effective theory. 

\nonfrenchspacing

\vskip 0.8cm
\vfill
\noindent $^1$ e-mail: {\em maarten@stars.sfsu.edu}  \\
\noindent $^2$ e-mail: {\em sharpe@phys.washington.edu}  \\
\noindent $^3$ e-mail: {\em bobs1@lanl.gov}  \\

\end{titlepage}
\section{Introduction}
\label{sec:intro}

Wilson fermions are one of the oldest and most important methods
for formulating the quark sector of QCD on a lattice \cite{wilson}.
Not only have they (or improved versions) been used directly for many numerical
computations, but they also play an important role in the more
recent domain-wall \cite{dk,ys} and overlap \cite{hn} formulations
of lattice QCD.

Wilson fermions do not preserve the chiral symmetry of continuum QCD,
and as a consequence the quark mass has to be tuned to recover
chiral symmetry in the continuum limit \cite{ks}.
Flavor and parity are exactly preserved, but, as observed by
Aoki~\cite{aoki}, these symmetries may be
spontaneously broken by a pionic condensate for certain values of
the bare quark mass.
This leads to the existence of the so-called Aoki phase.
In the theory with two flavors,
a pionic condensate (which we
can choose to point in the 3-direction in isospin space) breaks the
$SU(2)$ of isospin down to $U(1)$, giving rise to 
two Goldstone bosons at non-zero lattice spacing.
Within the Aoki phase, the third pion has a mass proportional to $a$,
the lattice spacing. 
In the continuum limit, the Aoki phase shrinks to zero width,
the pions all become massless, and the pionic condensate can
be rotated (by a non-singlet axial transformation) into the usual
condensate associated with spontaneous chiral symmetry breaking.  

The existence of the Aoki phase was investigated by two of 
us \cite{shsi} using the chiral effective lagrangian for pions 
in two-flavor unquenched lattice QCD (see also Ref.~\cite{creutz}).  
It was found that indeed an Aoki phase
may occur near the continuum limit, if a certain low-energy
constant in the effective lagrangian has a particular sign.
For the other choice of sign, however, there was no spontaneous
breakdown of flavor and parity, but rather a first order transition,
in the vicinity of which the three degenerate pions have masses
proportional to $a$.

In this paper we attempt to extend the effective lagrangian 
investigation of Ref.~\cite{shsi} to the quenched theory.
We have several motivations for doing so.
First, the issue is theoretically interesting, and challenging,
because of the presence of ghost-quarks in the lagrangian
formulation of
quenched QCD. While the peculiarities and pathologies of the
quenched theory do show up in perturbative investigations of
the effective theory~\cite{bg,sharpe}, they are more prominent
in the non-perturbative analysis needed to study the phase 
structure.\footnote{%
We recognize that the quenched theory is not a thermodynamic system in the
usual sense, as the quarks do not influence the gluon fields.
Nevertheless, the properties of quarks propagating on quenched gluon
fields as a function of the bare quark mass exhibit the non-analyticities
familiar from thermodynamics, so we think it appropriate to use
the term ``phase structure.''}

Second, the issue is of practical importance
for domain-wall and overlap fermions. Both are built
upon the (hermitian) Wilson-Dirac operator with
a large negative quark mass. This operator is effectively
 quenched (even in dynamical domain-wall or overlap simulations)
because its quark mass is not related to that of the physical
quarks. As explained in detail in Ref.~\cite{gsq},
the negative quark mass must be
chosen so that one does not lie in or near any  Aoki phase.
The arguments of Ref.~\cite{gsq} also imply
that, if there is a first order transition like that of the unquenched
theory, one should not simulate near the phase transition line.
The essential point is to avoid regions where the quenched pion correlators
are long ranged.
Thus it is clearly important to understand the phase structure
of the quenched theory.

Finally, numerical simulations provide some evidence
for the presence of the Aoki phase in the quenched theory~\cite{aokiq}.
Indeed, results from quenched simulations led Aoki to make his
original proposal of a new phase.\footnote{%
In the unquenched theory, recent simulations suggest, however, that,
with the Wilson gauge action, the scenario with a first-order
phase transition applies~\cite{desy}.}

More recent work has shown, however, that the situation in
the quenched theory is different from, and more subtle than,
that in the unquenched theory.
In particular, it was observed (see
for example Ref.~\cite{scri}) that there always appears to be
a non-vanishing density of near-zero modes of the (hermitian)
Wilson--Dirac operator in quenched QCD if the quark mass is
in the super-critical region.\footnote{%
Defined as the region
where the bare quark mass satisfies $-8r< ma < 0$.
This is where the Wilson--Dirac operator can,  in principle, have
exact zero modes.}  
This implies a non-vanishing pionic condensate, and thus,
through the Banks--Casher relation \cite{bc},
would seem to lead to the conclusion that an Aoki
phase fills the whole supercritical region,
in contrast to what is expected in unquenched QCD.  It was
shown in Ref.~\cite{gsq} that this is not the case, however,
if one defines the Aoki phase as that region of the phase
diagram where Goldstone bosons associated with the symmetry
breaking occur. To see this,  
Ref.~\cite{gsq} considered quenched QCD with two
flavors in the presence of a twisted quark mass, which 
explicitly breaks flavor and parity symmetry.
It was argued that (in the limit of vanishing twisted quark mass)
regions may exist where
the condensate does not vanish because of the existence of
a density of exponentially localized near-zero modes, 
without any of the corresponding long-range physics usually associated
with spontaneous symmetry breaking.  This phenomenon
is an artifact of the quenched approximation, and was shown to
be consistent with the usual Ward-identity argument for the
existence of Goldstone bosons.  It turns out that, in the quenched
case only, the Ward identity can be satisfied without a Goldstone
pole even in the presence of a non-vanishing condensate, if this
condensate arises because of localized near-zero modes \cite{gsq}.
Since the localization length of these near-zero modes is
of order the lattice spacing, this phenomenon leaves no trace
in the long-distance behavior of the quenched theory.

The implication of the analysis of Ref.~\cite{gsq} is that,
in the quenched theory, 
there is not a one-to-one relation between 
the pionic condensate in an effective field theory
(which is sensitive only to long-distance physics)
and that determined on the lattice. While a prediction of
a non-zero pionic condensate in the effective field theory implies
the presence of a lattice pionic condensate, 
the converse is not true. Thus numerical evidence of a condensate
is not, by itself, pertinent to the question of the phase structure.
What is pertinent, however, is the presence of a region in
the phase diagram in which there are long distance correlation lengths,
\ie\ pion masses satisfying $m_\pi \ll \Lambda_{\rm QCD}$.
An effective low-energy theory can be used in any such region.
A considerable body of numerical evidence indicates that the quenched
pion mass does extrapolate to very small values.
Based on this, we assume that there is a region  where an effective
theory can be used, and then study its properties.
 
To carry out such a study  we need a quenched effective lagrangian that 
is suited
to non-perturbative investigations.  A systematic approach to the quenched
theory along the lines of the continuum development of Ref.~\cite{gl} 
was given in Ref.~\cite{bg}.
It turns out, however, that, while this approach is 
sufficient for setting up chiral perturbation theory (ChPT), the
effective lagrangian given in~\cite{bg} is not suitable for the study 
of the phase
structure of the theory.

In the approach of Ref.~\cite{bg}, 
a ghost quark is introduced for every valence quark~\cite{morel}.
The ghost quarks couple to the gluons in the same way as the valence
quarks, and have the same mass, spin and flavor symmetries.  The only
difference is that ghost quarks have bosonic statistics.  Their
determinant cancels that from the valence quarks, thus providing a
path-integral definition of quenched QCD.  (For the extension to the case
with both valence and sea quarks, see Ref.~\cite{bgpq}.)
The ghost sector was dealt with only formally in
Ref.~\cite{bg}, without regard to the convergence of the ghost-quark
path integral.  Since the ghosts are bosonic, this is a non-trivial issue.
While the formal treatment of Ref.~\cite{bg} is sufficient to develop
quenched ChPT \cite{shsh}, a more careful treatment leads to a somewhat
different symmetry structure of the quenched QCD lagrangian
\cite{rm}. 
This different symmetry
structure leads to a different chiral lagrangian, which, while equivalent
to the one of Ref.~\cite{bg} for ChPT, 
is also suitable for non-perturbative
investigations.  The construction of this lagrangian and its use to
investigate the Aoki phase for the two-flavor theory are the 
central subjects of this paper.

We should stress at the outset that our analysis is, in several
respects, incomplete. At various stages we are forced to make 
additional assumptions not required in the corresponding analysis
of the unquenched theory. While we think our assumptions are
reasonable, it would clearly be preferable to avoid them.

The plan of the paper is as follows.
In Sect.~2 we define quenched QCD with $N$ flavors of Wilson fermions,
paying careful attention to the convergence of the ghost-quark 
path integral.  In order to do this, we formulate the theory in euclidean
space, as is done in numerical simulations. 
In Sect.~3 we discuss the symmetries of this theory in detail,
and then use these in Sect.~4
to construct the chiral effective lagrangian both 
in the continuum limit and including the leading effects
of discretization. In Sect.~5
we employ the resulting effective potential,
as well as certain general properties of the
quenched theory and the large-$N_c$ limit ($N_c$ is the number of colors),
to analyze the two-flavor theory.
We end with a summary and some concluding remarks.
An appendix briefly reviews illustrative calculations of
quenched small-volume partition functions.
A preliminary account of this work was given in Ref.~\cite{gsslat2004}.

\def\mmbf#1{\mbox{\boldmath${#1}$}}        
\section{$\mmbf{N}$-flavor Quenched QCD with Wilson Fermions}
\label{sec:setup}

In order to define quenched QCD, we introduce,
for every physical quark field $q$,
a ghost-quark field $\qt$ with the same
quantum numbers (spin, flavor and color)
as $q$ but opposite statistics~\cite{morel}.
If there are $N$ flavors, both $q$ and $\qt$ are $N$-dimensional vectors
in flavor space.
In the continuum, the euclidean lagrangian for quenched QCD
may then be defined as

\vbox{
\ba
\label{LCONT}
\cL&=&\qbar_L Dq_L+\qbar_R Dq_R+\qbar_R Mq_L+\qbar_L\Mbar q_R 
\\[3pt] \nonumber
&&+\qtd_R D\qt_L+\qtd_L D\qt_R+\qtd_L M\qt_L+\qtd_R\Mbar\qt_R \;,
\ea
}

\noindent
where

\vbox{
\ba
\label{PROJ}
&&q_L=P_Lq\,,\ \ \ q_R=P_Rq\,, \cr
&&\qbar_L=\qbar P_R\,,\ \ \ \qbar_R=\qbar P_L\,,
\ea
}

\noindent
in accordance with standard conventions, while,
in the ghost sector, 
\ba
\label{GHPROJ}
&&\qt_L=P_L\qt\,,\ \ \ \qt_R=P_R\qt\,, \cr
&&\qtd_L=\qtd P_L\,,\ \ \ \qtd_R=\qtd P_R\;.
\ea
Our convention is $P_{L,R} =(1\pm\gamma_5)/2$.
We take the mass matrices to satisfy $\Mbar=M^\dagger$;
numerical simulations usually involve real diagonal 
mass matrices, with $M=M^\dagger=\Mbar$.

In euclidean space, the Grassmann variables $q_{R,L}$ and
$\qbar_{L,R}$ are all independent, and the integration over them
leads to the quark partition function which is just the determinant
of the fermionic operator,
\be
\label{FDET}
\Det\left(D+MP_L+\Mbar P_R\right)\,.
\ee
This result holds for arbitrary mass matrices $M$ and $\Mbar$.
For the integral over the bosonic ghost fields to converge, however,
we must restrict $M$ and $\Mbar$ to be hermitian with all eigenvalues
positive. (The euclidean Dirac operator $D$ is anti-hermitian, and thus
plays no role in the convergence.)
Since we are taking $\Mbar=M^\dagger$, this means that $M=\Mbar$.
The gaussian integral then leads to the ghost-quark partition function
\be
\label{GDET}
\Det^{-1}\left(D+MP_L+\Mbar P_R\right)=\Det^{-1}\left(D+M\right)\,.
\ee
This cancels the quark determinant, and we see that indeed Eq.~(\ref{LCONT})
describes quenched QCD.  
We stress that $\qt_{R,L}^\dagger$ must be the 
hermitian conjugates of $q_{R,L}$ for the integral over
ghost fields to converge, while
no such connection exists between $q$ and $\qbar$.
These facts will have implications for the
symmetries of quenched QCD, to be discussed in the next section.

We would like to extend this continuum (and thus still formal)
definition to lattice QCD with Wilson fermions.  In the lattice action,
$D$ is replaced by the naive (nearest-neighbor) lattice discretization,
which retains the anti-hermiticity of the continuum $D$. 
The mass term $M$ is replaced by $M+W$, where $M$ is a local 
lattice mass term and $W$ the Wilson term,
\be
\label{WMASS}
(\qbar_R Wq_L)(x)=-\frac{1}{2}\sum_\mu\qbar_R(x) r \left(U_\mu(x)q_L(x+\mu)+
U^\dagger_\mu(x-\mu)q_L(x-\mu)-2q_L(x)\right)\,.
\ee
The corresponding term with $L\leftrightarrow R$,
which we call $\overline W$, contains $\bar r$.
In general we need to take $r$ and $\bar r$ to be flavor matrices,
so that we can treat them as spurion fields.
They will always be related by $\bar r = r^\dagger$.
In simulations, however, one usually takes $r=\bar r$ to be the 
identity matrix,
and this is the choice we make for the remainder of this section.

As is well known, in order to approach
the chiral limit at non-zero lattice spacing one must work
at negative bare quark mass. This is necessary because
$W$, being positive semi-definite, makes a positive contribution to
the physical quark mass. But once the quark mass is negative,
so that $M$ has negative eigenvalues, there are some configurations on which
$D+M+W$ itself can have eigenvalues with a negative or vanishing real part.
(The case of vanishing real part corresponds to the so-called
``exceptional configurations.'')
It follows that we cannot simply take over the definition (\ref{LCONT}),
with $M\rightarrow M+W$, to define the quenched lattice theory,
since the ghost-quark integral would be ill-defined.
The same holds true in the continuum if $M$ has negative eigenvalues.

In these cases we must proceed in a different way.  
We begin with the unquenched quark lagrangian for Wilson fermions
with $M=\Mbar$, and we work in a basis in which the mass matrix 
$M$ is diagonal, while adding an infinitesimal parity-odd 
mass term:
\be
\label{POMASS}
\cL_{\rm quark} = \sum_{j=1}^N \Big[\,
\qbar'_j (D + M_j + W) q'_j
- \epsilon_j\; \qbar'_j i \gamma_5 q'_j\,\Big]
\,.
\ee
Here $j$ is the flavor index, and the infinitesimal parameters $\epsilon_j$
can have either sign independently for each flavor. 
We use primed fields in Eq.~(\ref{POMASS}) because, in order to avoid the
problem of non-positive eigenvalues of $D+M+W$, we perform a change
of variables given by the following axial transformation:
\be
\label{TRANSF}
q'_j = \Exp\left[{\rm sgn}(\epsilon_j) i\frac{\pi}{4}\gamma_5\right]\;q_j\,,
\ \ \ \ 
\qbar'_j = \qbar_j\;\Exp\left[{\rm sgn}(\epsilon_j) i\frac{\pi}{4}
\gamma_5\right]\,.
\ee
The quark-sector lagrangian becomes
\be
\label{QUARK}
\cL_{\rm quark}=\sum_{j=1}^N  \Big[\,\qbar_j\Big(D + {\rm sgn}(\epsilon_j)\; 
i\gamma_5(M_j + W)\Big)q_j + |\epsilon_j| \;\qbar_j q_j \,\Big]  \,.
\ee
The resulting
lattice Dirac operator $D\pm i\gamma_5(M+W)$ is anti-hermitian,\footnote{%
This form is unfamiliar, but is, in fact,
unitarily equivalent to $i$ times the well-known
hermitian Wilson-Dirac operator, $ H_W=\gamma_5 [D\pm(M+W)]$.}
and thus is suitable for extension to the ghost sector.
Furthermore, the $\epsilon_j$ terms (which, after the axial transformation,
look like regular mass terms) ensure convergence once we extend
to the ghost sector. 
Note that for this to be true the direction of the axial rotation in
Eq.~(\ref{TRANSF}) must be correlated with the sign of the infinitesimal parity-odd
mass term in Eq.~(\ref{POMASS}).

We can now define quenched lattice QCD with Wilson fermions
by adding the ghost sector:
\ba
\label{LLATT}
\nonumber
\cL_W  &=&\sum_{j=1}^N \Big[\,\qbar_j\Big(D+ {\rm sgn}(\epsilon_j) 
i\gamma_5(M_j + W) \Big)q_j + 
\qtd_j\Big(D+ {\rm sgn}(\epsilon_j) i\gamma_5(M_j + W)
\Big)\qt_j
\\
&& \hskip1cm  + ~|\epsilon_j| (\qbar_j q_j + \qtd_j\qt_j)
\, \Big] \,.
\ea
Because of the exceptional configurations mentioned above,
quenched correlation functions diverge, in general,
in the limit $\epsilon_j\to 0$, since zero modes of
$D+M+W$ are also, after the axial rotation, zero modes of
$D \pm i\gamma_5(M+W)$\,\cite{bardeen}.  We 
therefore keep the $\epsilon_j$ non-zero, but infinitesimal.

Several comments are in order.
First, one may worry that the transformation of Eq.~(\ref{TRANSF}) 
could be anomalous.  This is not the
case.  The fermionic measure on the lattice is rigorously 
invariant, and no vacuum angle $\theta$ is generated.
This is in accordance with the fact that in order to produce a
non-zero vacuum angle from Wilson fermions, one has to introduce
a relative phase between the Wilson mass ($W$) and the single-site
mass ($M$) \cite{sest}.\footnote{See Ref.~\cite{gs} for an alternative method.}

Second, the fact that the sign of $\epsilon_j$ may be chosen independently for
each quark flavor has interesting consequences. For example, with two 
quenched flavors, one may choose $\epsilon\equiv\epsilon_u=-\epsilon_d>0$
for the up and down flavors.
This corresponds to (quenched) twisted-mass lattice QCD~\cite{aoki,tm}.  
It leads, after the axial transformation, to the lagrangian
\be
\label{QTWOFLAVOR}
\cL_W^{\rm 2-flavor}=\qbar\Big(D+i\tau_3\gamma_5(M+W)\Big)q +
\qtd\Big(D+i\tau_3\gamma_5(M+W)\Big)\qt+\epsilon (\qbar q + \qtd\qt)\,,
\ee
where $\tau_3$ is the diagonal Pauli matrix acting in flavor space.
We see that, in order to guarantee convergence of the ghost-quark
integral, a non-trivial flavor dependence is introduced into
the quark-mass and Wilson terms.  If we choose $\epsilon_u$ and
$\epsilon_d$ of the same sign, however, no such flavor dependence appears.
We will analyze both cases in this paper.

For later use, we rewrite the lagrangian (\ref{LLATT}) in two other ways.
For the sake of clarity we give the expressions only for
the simplest choice of the $\epsilon_j$, namely that all are
equal and positive: $\epsilon_j=\epsilon >0$. 
It is straightforward to generalize the expressions to any other choice
for the $\epsilon_j$ by going back to Eq.~(\ref{LLATT}).
We first rewrite ${\cal L}_W$ in terms of the $\gamma_5$ projected fields of 
Eqs.~(\ref{PROJ},\ref{GHPROJ}):
\ba
\label{LLATTPROJ}
\cL_W&=&\qbar_L Dq_L+\qbar_R Dq_R+\qbar_R\Mt q_L+
\qbar_L\Mt^\dagger q_R
\\[3pt] \nonumber
&&+\qtd_R D\qt_L+\qtd_L D\qt_R+\qtd_L\Mt\qt_L+\qtd_R\Mt^\dagger\qt_R\,,
\ea
with $\Mt=i(M+W-i\epsilon)$, and $\epsilon$ implicitly multiplied by the
identity matrix in flavor space. 
Next, we group the quark and ghost-quark fields into a ``super field"
\be 
\label{PSI}
\Psi_{L,R}=\pmatrix{q_{L,R}\cr\qt_{L,R}}\,,\ \ \ \ 
\Psibar_{R,L}=\pmatrix{\qbar_{R,L}&\qtd_{L,R}}\,,
\ee
in terms of which
\be
\label{LSUPER}
\cL_W=\Psibar_LD\Psi_L+\Psibar_RD\Psi_R+\Psibar_R\cM\Psi_L
+\Psibar_L\cMbar\Psi_R\,,
\ee
with 
\be
\label{MASSMATRIX}
\cM=i\pmatrix{M+W - i\epsilon&0\cr 0&M+W-i\epsilon}\,,
\ee
and $\cMbar=\cM^\dagger$.

\section{Symmetries of Continuum Quenched QCD}
\label{sec:symm}

In this section we consider the symmetries of quenched QCD in the
continuum limit. In that limit, the Wilson term is irrelevant once the
additive renormalization of the quark mass has been included. Thus we
can treat the quantity $M+W$ as simply a quark mass matrix.
Note that, because of our use of the axial transformation (\ref{TRANSF}),
we can treat the quenched theory for either sign of the quark masses.

It is instructive to first discuss the quark
and ghost sectors separately.  For $\Mt=0$, the
quark part of $\cL_W$ is invariant under
\be
\label{QSYMM}
q_{L,R}\to V_{L,R}\,q_{L,R}\,,\ \ \ \ 
\qbar_{L,R}\to\qbar_{L,R}\,V^{-1}_{L,R}\,,
\ee
with $V_L$ and $V_R$ both elements of $GL(N)$ (for $N$ flavors).
Thus the chiral symmetry group is $GL(N)_L\times GL(N)_R$.
(This ignores the anomaly, to which we return below.)
This can be maintained as a symmetry of the full quark lagrangian 
if $\Mt$ and $\Mt^\dagger$ are treated as {\em independent}
spurion fields. Renaming the latter as $\Mt^\dagger\to\Mtb$,
the required transformations are 
\be
\label{MTRANSF}
\Mt\to V_R \Mt V^{-1}_L\,,\qquad
\Mtb \to V_L \Mtb V^{-1}_R \,.
\ee
We take the vector subgroup to be that which remains when
$M$ and $r$ are proportional to the identity matrix and all 
$\epsilon_j$ are equal.
This requires $V_L= V_R$,
so that the vector subgroup is the diagonal subgroup $GL(N)$.
  
The symmetry group of the quark sector in the euclidean
formulation is larger than that of the hamiltonian formulation, where
the group $GL(N)$ is reduced to the unitary group $U(N)$.  It is 
straightforward to show that the larger group (which is just the
complexification of $U(N)$) does not lead to any new Ward identities
(see for instance Ref.~\cite{shsh}).

In the ghost sector, the symmetry transformations of 
$\qt_L$ and $\qt_R$ are coupled, and, setting $\Mt=0$ 
in Eq.~(\ref{LLATTPROJ}), the ghost lagrangian is 
invariant under \cite{rm}
\be
\label{GHSYMM}
\qt_L\to V\qt_L\,,\ \ \ \ \qt_R\to V^\dagger{}^{-1}\qt_R\,,
\ee
with $V\in GL(N)$.  In other words, if $\qt_{L,R}\to V_{L,R}\qt_{L,R}$,
the form of the ghost lagrangian leads to the requirement that
\be
\label{GHRESTR}
V_L=V_R^\dagger{}^{-1}\equiv V\,.
\ee
The transformations of the spurion fields $\Mt$ and $\Mtb$ are
\be
\label{GHSPURION}
\Mt\to V^\dagger{}^{-1}\Mt V^{-1}\,,\qquad 
\Mtb\to V\Mtb V^\dagger\,.
\ee

The vector subgroup, which we define as the subgroup which
leaves $\qtd_L \qt_L$ and $\qtd_R \qt_R$
invariant, has $V^\dagger V=1$, \ie\ $V\in U(N)$.
If we parametrize $V\in GL(N)$ as
\be
\label{VPAR}
V=\Exp(\alpha_0+ \sum_i \alpha_i T_i)\,,
\ee
with $T_i$ the hermitian generators of $SU(N)$ 
and $\alpha_{0,i}$ complex,
then the vector subgroup has $\alpha_{0,i}$ imaginary.
``Axial'' transformations are those with
 $\alpha$ and $\alpha_i$ real, so that $V=V^\dagger$.

\bigskip

Next, we discuss the symmetries of the full lagrangian, considering also
transformations of quarks into ghost quarks and {\it vice versa}.
For this, it is useful to start from the form of the lagrangian
given in Eq.~(\ref{LSUPER}). If $\Psi$ and $\Psibar$ were
independent, then $\cL_W$ would be invariant under
\be
\label{SFTRANSF}
\Psi_{L,R}\to\cV_{L,R}\Psi_{L,R}\,,\ \ \ \ 
\Psibar_{L,R}\to\Psibar_{L,R}\cV^{-1}_{L,R}\,,
\ee
with $\cV_{L,R}\in GL(N|N)$, as long as we
treat $\cM$ and $\cMbar$ as independent spurion fields transforming as
\be
\label{SPURION}
\cM\to\cV_R\cM\cV_L^{-1}\,,\ \ \ \cMbar\to\cV_L\cMbar\cV_R^{-1}\,.
\ee
This would imply that the (massless) quenched theory possesses the
graded chiral symmetry group $GL(N|N)_L\times GL(N|N)_R$.
This does not take into account, however,
 that $\Psibar_{L,R}$, when restricted
to the ghost sector, is not independent of $\Psi_{L,R}$, but rather,
that
\be
\label{RESTR}
\Psibar_{L,R}|_{\rm ghost,body}=\Psi^\dagger_{R,L}|_{\rm ghost,body}\,.
\ee
What this equation means is the following.
After a graded transformation, the ghost field $\qt$ picks
up terms containing Grassmann numbers. 
In general, $\qt$ can be split into its ``body" and a 
``soul," where the body is the usual complex field, and the soul
is even in Grassmann fields or parameters.  As discussed in more
detail in Ref.~\cite{shsh}, it is sufficient for the convergence
of the ghost integral that Eq.~(\ref{RESTR}) apply only to the
body of the ghost fields, as already indicated in that equation.
It can then be shown that the restriction of Eq.~(\ref{GHRESTR})
generalizes to \cite{shsh}
\be
\label{GRRESTR}
\cV_{Lgg}|_{\rm body}=\cV_{Rgg}^{\dagger-1}|_{\rm body}\,,
\ee
where, in $N\times N$ block form (for $N$ flavors),
\be
\label{GHPART}
\cV_L=\pmatrix{\cV_{Lqq}&\cV_{Lqg}\cr\cV_{Lgq}&\cV_{Lgg}}\,,
\ee
and similar for $\cV_R$.  The label $q$ refers to the quark sector,
and the label $g$ to the ghost sector.  Note that the resulting
set of transformations, 
\be
\label{GROUPWITHANOMALY}
\cG'=\Big\{(\cV_L,\cV_R)\in GL(N|N)_L\times GL(N|N)_R\ 
~\Big\vert ~ \cV_{Lgg}|_{\rm body}=\cV_{Rgg}^{\dagger-1}|_{\rm body}
\Big\}\,,
\ee
still forms a group.

To complete the discussion of continuous symmetries we review the
impact of the anomaly~\cite{bg}. In the continuum limit,
transformations in $\cG$ for which
$\sdet(\cV_L) \ne \sdet(\cV_R)$ are anomalous,
and should be excluded.
This can be accomplished by
restricting $\cV_{L,R}$ to lie in $SL(N|N)$ rather than
$GL(N|N)$. There is one subtlety, however.
The non-anomalous vector $U(1)$ transformations do not commute
with general elements of $SL(N|N)$ and so the complete
symmetry group is a semi-direct product \cite{bglat92}:
\be
\label{GROUP}
\cG=\Big\{(\cV_L,\cV_R)\in
[SL(N|N)_L\times SL(N|N)_R]\ltimes U(1)_V ~ \Big\vert ~ 
\cV_{Lgg}|_{\rm body}=\cV_{Rgg}^{\dagger-1}|_{\rm body}
\Big\}\,.
\ee
The vector subgroup (the subgroup which leaves $\Psibar \Psi$ invariant)
is not affected by the anomaly and
is given by
\be
\label{FULLVECTOR}
\cH=\Big\{(\cV=\cV_L=\cV_R)\in GL(N|N) ~ \Big\vert ~
\cV_{gg}|_{\rm body}=\cV_{gg}^{\dagger-1}|_{\rm body}
\Big\}\,.
\ee
For further discussion of the
symmetry group, see Ref.~\cite{shsh}.

\bigskip
Finally, we consider the properties of $\cL_W$
under parity.  Unquenched QCD with Wilson fermions in the standard
form is invariant under parity, but this is not obviously true for
our definition of quenched QCD.  In terms of the new quark
variables of Eq.~(\ref{TRANSF}), a parity transformation acts on
the spinor indices as
\be
\label{PARITY}
q(t,\vx)\to i\gamma_4\gamma_5 q(t,-\vx)\,,\ \ \ \ 
\qbar(t,\vx)\to\qbar(t,-\vx) i\gamma_5\gamma_4\,,
\ee
and the quark sector of $\cL_W$ is invariant under this transformation
(if the gauge field is transformed accordingly).  The ghost
sector is not, however, invariant under this symmetry.  
If we try to define parity on the ghost field as
\be
\label{GHPARITY}
\qt(t,\vx)\to i\gamma_4\gamma_5\qt(t,-\vx) 
\Rightarrow \qtd(t,\vx)\to-\qtd(t,-\vx) i\gamma_5\gamma_4\,,
\ee
the ghost part of $\cL_W$ transforms into minus itself.  

However, as long as we consider correlation functions involving only
physical quark fields parity is not broken.  Parity 
breaking can then only come from ghost loops, but these are always 
exactly cancelled by physical quark loops.  A more formal version of
the argument follows when we couple only the physical quarks to
external sources.  After integrating over the quark and ghost fields,
the determinants cancel, and only the physical quark propagator
couples to the sources.  We conclude that parity is not broken in
quenched QCD, with the proviso that we only consider correlation
functions made out of physical quark fields.

It will be useful to consider also ``naive" parity:
\ba
q(t,\vx)&\to&\gamma_4 q(t,-\vx)\,,\ \ \  
\qbar(t,\vx)\to\qbar(t,-\vx)\gamma_4\,,\label{NAIVEP}\\
\qt(t,\vx)&\to&\gamma_4\qt(t,-\vx)\,,\ \ \ 
\qt(t,\vx)^\dagger\to\qt(t,-\vx)^\dagger\gamma_4\,,\nonumber\ea
Under this transformation,
the kinetic terms of both the quark and ghost
sectors are invariant, but not the mass terms, which transform
into minus themselves because of the $\gamma_5$ they contain.
The lagrangian (\ref{LSUPER}) is, however,
 invariant if we transform the
spurion fields $\cM$ and $\cMbar$ under naive parity as
\be
\label{NPMASS}
\cM\to\cMbar\,,\ \ \ \cMbar\to\cM\,.
\ee

\section{Quenched Chiral Effective Lagrangian}
\label{sec:leff}

In this section, we discuss the construction of the effective theory
for the Goldstone particles of quenched QCD, assuming that the chiral 
symmetry is spontaneously broken 
and that Goldstone-like  excitations occur.
These assumptions are based mainly on the results
from numerical simulations. The order parameter for symmetry breaking is
the quark condensate, which, for positive degenerate quark masses, 
is aligned so that the vector symmetry is unbroken.

We will need to address several issues that arise in the quenched
theory that are not present when constructing the effective theory
for unquenched QCD. The first is the fact that, in the quenched theory,
a non-vanishing condensate does not necessarily lead to Goldstone-like
excitations~\cite{gsq}. The Ward identity can be saturated by
localized near-zero modes of the lattice Dirac operator.  
According to the conjecture of Ref.~\cite{gsq},
such near-zero modes are present throughout the supercritical region,
except where there are extended near-zero modes. In the latter case,
the Ward identity is saturated by the usual Goldstone particles.
The net effect of these observations is that the condensate in the
chiral lagrangian cannot be identified with that at the quark level,
but rather with a ``long-distance" condensate defined by removing
the contributions of short-distance near-zero modes.  Since the
$\epsilon_j\to 0$ divergences in quenched correlation functions are
due to localized near-zero modes \cite{gsq}, we conjecture that
the limit $\epsilon_j\to 0$ can be taken after the removal of these
short-distance near-zero modes.\footnote{In principle, this can be
done by allowing only admissible gauge fields \cite{hjl}.}

The second issue concerns the presence of infra-red divergences
in the chiral limit which make this limit singular~\cite{bg,sharpe}.
To avoid these, we assume that there is a region with small enough
(but non-zero) quark masses for which the correlation lengths of
Goldstone correlation functions are much larger than those of
other excitations in the theory, so that it makes
sense to use the effective lagrangian approach.  

The third issue is the need to keep the ``singlet'' Goldstone field,
$\Phi_0$ (defined precisely below),
despite the fact that the corresponding symmetry is anomalous.
This point is explained in Ref.~\cite{bg,shsh,sharpe}:
$\Phi_0$ must be kept in the effective theory because
its correlators have long distance contributions, even though
some of these contributions are not those of a standard 
single-particle pole.

The remaining issues will be discussed as they arise in the
rest of this section.
In the first subsection
we discuss the effective theory for quenched QCD in the continuum
limit.
In the second subsection we extend this to include terms of order $a$
and $a^2$, where $a$ is the lattice spacing.

\subsection{Continuum Effective Lagrangian}

Here we construct the effective lagrangian for
quenched QCD in the continuum limit, starting from the formulation
of quenched QCD developed in Sect.~2.  This was done before in 
Ref.~\cite{bg}, but there it was naively assumed that the full
chiral symmetry group is the graded group $U(N|N)_L\times U(N|N)_R$,
an assumption which we have seen above not to be entirely correct.
It turns out that the effective lagrangian is equivalent to 
that of Ref.~\cite{bg} if the aim is only to develop chiral
perturbation theory for quenched QCD, but not if one wants to
do non-perturbative calculations, such as needed to explore
the Aoki phase.

Following the standard development, we expect the 
Goldstone excitations to be described by a non-linear field
$\Sigma$ which transforms as $\Psi_L\Psibar_R$ and spans
the coset $\cG/\cH$:
\be
\label{SIGMA}
\Sigma=\Exp(\Phi)\,,\quad
\Sigma\to\cV_L\Sigma\cV_R^{-1}\,.
\ee
Here $\cV_{L,R}$ are elements of the symmetry group $\cG$,
Eq.~(\ref{GROUP}), and $\cH$ is the unbroken subgroup of 
Eq.~(\ref{FULLVECTOR}).
If $\langle \Sigma\rangle$ is proportional
to the identity matrix, then 
$\Phi$ is a linear combination of the broken generators.
To construct a lagrangian invariant under $\cG$, we also need
 $\Sigma^{-1}\to\cV_R\Sigma^{-1}\cV_L^{-1}$, 
transforming as $\Psi_R\Psibar_L$. 
As already noted above, however, we must enlarge $\Sigma$ to include
the ``super-$\eta'$" field
$\Phi_0\equiv -i\,\str\log\Sigma=-i\;\log\sdet\Sigma$
($\str$ is the supertrace, $\sdet$ the superdeterminant).
This can be done by allowing $\cV_{L,R}$ to be elements
of the larger, anomalous, group $\cG'$, Eq.~(\ref{GROUPWITHANOMALY}),
while still only requiring the lagrangian to be invariant
under $\cG$. Since $\Phi_0$ is invariant under $\cG$, this leads
to the usual presence of an arbitrary set of functions of this singlet
field in the effective lagrangian \cite{gl,bg}. 

With these ingredients, and treating the masses as
spurions as described above, the $O(p^2)$ 
effective lagrangian becomes 
\ba
\cL_{\rm eff}&=&\frac{1}{8}f^2 V_1(\Phi_0)\;
\str(\partial_\mu\Sigma\partial_\mu\Sigma^{-1})
-v V_2(\Phi_0)\;\str(\cM\Sigma+\Sigma^{-1}\cMbar) \label{LEFFANOM}\\
&&+\frac{1}{2}c_0\Phi_0^2 V_0(\Phi_0)+
\frac{1}{2}\alpha(\partial_\mu\Phi_0)^2 V_5(\Phi_0)\,,\nonumber
\ea
with $V_i(\Phi_0)=1+O(\Phi_0^2)$, $i=0,1,2,5$.
We have used a field redefinition to remove possible
terms linear in $\Phi_0$ \cite{gl,bg}.
The constant $c_0$ is of $O(\Lambda^2)$ ($\Lambda\equiv \Lambda_{\rm QCD}$ 
being the non-perturbative scale of QCD) and, because
of its relation to the $\eta'$ mass and topological susceptibility,
is expected to be positive~\cite{bg,sharpe}.
The mass matrix $\cM$, Eq.~(\ref{MASSMATRIX}),
becomes (taking all $\epsilon_j$ positive and equal for simplicity)
$\cM=i(m-i\epsilon)$ with $m= M-m_c(r)$ the subtracted mass matrix,%
\footnote{$m_c(r)$ is the critical quark mass.  As one approaches the
continuum limit, this can be defined as the value for which the
pions become massless. This definition does not work away from the
continuum limit, however, since the
subsequent analysis shows that the pions may have a minimal
mass of $O(a)$, or there may be a region of quark masses for
which there are massless pions. For our purposes, however, it is
sufficient to have a definition which determines $m_c$
to within an accuracy of $O(a\Lambda^2)$, and one such definition
is that the pion mass should be of size $a\Lambda^2$.}
and $\cMbar=\cM^\dagger$.
The plus sign in the mass term follows from the fact that under
naive parity $\Sigma\to\Sigma^{-1}$ and $\cM\leftrightarrow\cMbar$,
\cf\ Eq.~(\ref{NPMASS}). Note that, while this looks like the
standard mass term in ChPT, there are factors of $\pm i$ hidden
in $\cM$ and $\cMbar$, related to 
the original singlet axial field redefinition of Eq.~(\ref{TRANSF}).

We now consider how to parameterize the field $\Phi$.
Were it not for the restrictions on the ghost-ghost part of the
transformations, $\Phi$ would generate $GL(N|N)$ and thus be
a general complex graded matrix. The restriction (\ref{GHRESTR}) implies,
however, that the ghost-ghost part of $\Phi$ must have an hermitian
body, and so we parameterize it as
\ba
\label{PHIPAR}
\Phi=\left(
\begin{array}{cc}
 i\phi_1+\phi_2 & \chibar \\[3pt] 
\chi & \phibar
\end{array}\right)\,,
\ea
with 
$\phi_{1,2}$ hermitian $N\times N$ matrices of c-numbers, $\chi$ and
$\chibar$ $N\times N$ matrices of
independent Grassmann variables, and $\phibar$ an hermitian
c-number $N\times N$ matrix.  Note that, in addition, $\phi_{1,2}$ as well as
$\phibar$ can have arbitrary souls.

There are a number of unusual features of the parameterization
(\ref{PHIPAR}). First and foremost, the
presence of both $\phi_1$ and $\phi_2$ in Eq.~(\ref{PHIPAR}) 
implies a redundancy in the number
of fields describing the physical pions.  
This redundancy is not, however, special to the quenched theory --- it
occurs also in unquenched QCD. The standard prescription is to set
$\phi_2=0$, so that $\Sigma$ is unitary in the quark sector, reflecting
the fact that only the unitary subgroup are symmetries of the hamiltonian.
We will follow this prescription here, and briefly review the arguments
supporting this choice for the quenched theory.\footnote{%
For a parallel discussion in the partially quenched case see Ref.~\cite{shsh}.}

First, we note that
the effective lagrangian depends only on the field
$\phi=\phi_1-i\phi_2$, and not on $\phi_1+i\phi_2$.  
If one is only
interested in developing chiral perturbation theory one may just expand
in $\phi$ (as well as $\phibar$, $\chi$ and $\chibar$). This leads to
the standard Feynman rules in the physical meson sector.\footnote{%
In fact, we are assuming implicitly in this discussion that we should
use the lagrangian (\protect\ref{LEFFANOM}) as is, rather than taking
its real part. We are also assuming that the parameters $f$, $v$, etc.
are real.}
However, if one wishes to consider
non-perturbative issues such as the phase diagram of the theory, or the
exact finite-volume partition function, the contour of integration of
$\phi$ needs to be specified.

We note in passing that the absence of a factor of $i$
multiplying $\phibar$ in (\ref{PHIPAR}) means that the propagator for
ghost-ghost mesons has the sign of a physical meson propagator
(due to the minus sign from the supertrace). 
This differs, superficially, from the standard Feynman
rules of quenched
chiral perturbation theory in which the ghost-ghost mesons have wrong
sign propagators~\cite{bg}. There are also, however, additional factors
of $i$ in vertices, and it is straightforward to see that these factors
can be shuffled from vertices to propagators in such a way as to
reproduce the standard rules. This is why the use of the
correct symmetry group is unnecessary when only developing
perturbation theory.
Of course, in order to see the relation to
standard quenched chiral perturbation theory one
must also ``undo" axial rotations of Eq.~(\ref{TRANSF}).

A stronger argument for setting $\phi_2=0$ is
that, if we choose to integrate the field $\phi$
along the contour $\phi_2=0$, the quenched small-volume partition
function is independent of the quark mass, as it should
be \cite{zirn,jv}.%
\footnote{Here ``small volume" refers
to the regime $m_\pi L\ll 1 \ll \Lambda_{\rm QCD}L$, with $L$
the linear size of the volume.}  A different contour would not
lead to the same result. In the Appendix we give a demonstration
of this in the one-flavor theory in the sector with zero
topological charge, using a parameterization which
makes the calculation particularly simple.  

Another feature of the parameterization (\ref{PHIPAR}) is the presence
of the ``souls" (nilpotent parts) of the fields $\phi_1$, $\phi_2$
and $\phibar$. In particular, the soul of $\phibar$ need not be hermitian,
although the body must be. As discussed in Ref.~\cite{shsh}, however,
the souls do not contribute to integrals, as long as the integrals are convergent.
Thus, in effect, we can treat $\phi_1$, $\phi_2$ and $\phibar$ as 
standard hermitian matrices, with the prescription discussed above
setting $\phi_2=0$.

One might be concerned that if we restrict the degrees of freedom
in $\Sigma$ by setting $\phi_2=0$, and thus reduce the set of allowed
transformations of $\Sigma$, we might have to allow additional terms to
appear in the effective lagrangian. We suspect that, in fact, the correct
approach is to use the full symmetries to determine the effective
lagrangian, and then restrict the manifold of $\Sigma$.
If so, this would remove the concern.
But, in any case, the remaining transformations are sufficient
to forbid additional terms. These transformations are unitary
in the quark-quark block, hermitian in the ghost-ghost block,
and arbitrary in the quark-ghost blocks.

\subsection{Order $\mmbf{a}$ Effects}

In this section, we consider how the flavor-symmetry structure of
quenched QCD with Wilson fermions is modified at non-vanishing lattice
spacing.  From now on, we will always choose
all physical and ghost quark masses equal to $m$.  In the unquenched case,
it was argued in Ref.~\cite{shsi} that the relevant continuum
quark lagrangian including terms of order $a$ (the lattice spacing)
is given by
\be
\label{LEFFQUARK}
\cL_{\rm quark,\,eff}=\cL_{\rm gluons}+\qbar(D+i\gamma_5 m)q+ a\qbar
b_1 i\gamma_5 \, i\sigma_{\mu\nu}F_{\mu\nu}q+O(a^2)\,,
\ee
with $b_1$ a coefficient which depends on the QCD coupling
constant.  The term linear in $a$ is the Pauli term.
The matrix $i\gamma_5$ appears in both the mass and
Pauli terms because of the field redefinition Eq.~(\ref{TRANSF}).
The Pauli term breaks
chiral symmetry in the same way as the mass term, and this 
symmetry breaking may thus be represented by a spurion field
$A$ which transforms just as $\Mt$ in Eq.~(\ref{MTRANSF}).

In the quenched case, analogous arguments lead to the quark
effective lagrangian
\be
\label{QLEFFQUARK}
\cL^{\rm quenched}_{\rm quark,\,eff}=\cL_{\rm gluons}+
\Psibar(D+i\gamma_5 m)\Psi+ a\Psibar b_1 i\gamma_5 \,
i\sigma_{\mu\nu}F_{\mu\nu}\Psi+\epsilon\Psibar\Psi+O(a^2)\,,
\ee
where $b_1$ can take a different value in the quenched case.
We have included the convergence term, taking
$\epsilon_j=\epsilon > 0$ for all flavors. Note that the 
effective Dirac operator $D+i\gamma_5 m-b_1 a\gamma_5
\sigma_{\mu\nu}F_{\mu\nu}$ is anti-hermitian,
as is the corresponding operator in the underlying lattice theory.

The extra $i\gamma_5$ appears in the mass and Pauli terms
because of the unusual form
(\ref{PARITY}) that parity takes in our formulation of quenched QCD.
In fact, the situation is somewhat subtle.  Parity is broken in
the ghost sector (\cf\ Sect.~3), and indeed, the ghost part
of the Pauli term as well as that of the mass term
in Eq.~(\ref{QLEFFQUARK}) breaks parity: like
the rest of the ghost lagrangian, 
it changes sign under (\ref{GHPARITY}).  However, as already
observed at the end of Sect.~3, if we restrict ourselves to
consider only correlation functions involving physical quark
fields, parity is not broken by the underlying lattice theory,
and correspondingly also not by $\cL^{\rm quenched}_{\rm quark,\,eff}$.
A term of the form $\Psibar\sigma_{\mu\nu}F_{\mu\nu}\Psi$
{\it cannot} appear, because it would break parity in the
physical sector.  Under naive parity (\cf\ Eq.~(\ref{NAIVEP}))
the Pauli terms also switch sign, just like the mass terms.

It follows that, as in the unquenched theory, the Pauli term breaks chiral symmetry
in the same way as the mass term.  We may introduce spurion
fields $\cA=ib_1$, $\cAbar=-ib_1=\cA^\dagger$ transforming
just like $\cM$ and $\cMbar$ in Eq.~(\ref{SPURION}) to keep
track of this symmetry breaking at the level of the
Goldstone-meson effective lagrangian.

With this new spurion field, we can extend the effective
lagrangian of Eq.~(\ref{LEFFANOM}) to include $O(a)$ effects.
The rule is simply that every appearance of $\cM$ could be 
replaced by $\cA$. We are interested in vacuum structure
and so will consider only the potential.
We keep all terms of size $m^2$, $ma$
and $a^2$ --- $O(p^4)$ in the usual chiral counting.
Recalling that all masses are degenerate, we find
\ba
\label{POTENTIAL}
V(\Sigma)&\!=\!&-ic_1\str(\Sigma-\Sigma^{-1})
-\epsilon\;\str(\Sigma+\Sigma^{-1})
+\frac{1}{2}\;c_0\Phi_0^2\\
&&+c_2\left((\str\Sigma)^2+(\str\Sigma^{-1})^2\right)
+c_3(\str\Sigma)(\str\Sigma^{-1})
+c_4\left(\str(\Sigma^2)+\str(\Sigma^{-2})\right)\,,\nonumber
\ea
with
\ba
\label{SOURCE}
c_1&=&\alpha_1 m\Lambda^3+\alpha_2 a\Lambda^5\,,\\
c_2&=&\beta_1 m^2\Lambda^2+\beta_2 ma\Lambda^4+\beta_3
a^2\Lambda^6\,,\nonumber\\
c_3&=&\gamma_1 m^2\Lambda^2+\gamma_2 ma\Lambda^4+\gamma_3
a^2\Lambda^6\,,\nonumber\\
c_4&=&\delta_1 m^2\Lambda^2+\delta_2 ma\Lambda^4+\delta_3
a^2\Lambda^6\,,\nonumber
\ea
Here $\alpha_i$, $\beta_i$, $\gamma_i$ and $\delta_i$ 
are dimensionless constants of order one,
and the factors of $\Lambda$ are as
required by dimensional analysis.  The $c_i$ terms 
in (\ref{POTENTIAL}) can
also be multiplied by $\Phi_0$ dependent potentials, 
but we do not make this explicit 
since we will argue in the next section that we can set $\Phi_0=0$
in the vacuum.

There is an apparent inconsistency in our analysis. We have kept
$O(a^2)$ terms in the effective lagrangian, but not in the underlying
quark lagrangian, $\cL^{\rm quenched}_{\rm quark,\,eff}$.  The missing terms
(\eg\ four-fermion operators)
do not,
however, break any further symmetries and so do not lead to
additional terms in $V(\Sigma)$. The only exception is a 
three derivative term $\sim \Psibar \gamma_\mu D_\mu^3 \Psi$,
which breaks the rotation group down to its hypercubic subgroup,
but this maps into a term with four derivatives in the effective
chiral lagrangian, and is thus absent from $V(\Sigma)$.

Finally, we address the issue of the relative size
of the coefficient $c_1$ and the coefficients $c_{2,3,4}$.
Following Ref.~\cite{shsi}, we distinguish three regimes:
\begin{itemize}
\item[1.]  The quark mass is physical while $a\to 0$.  This
means $m/\Lambda$ is a (small) constant, so that, for $a\to 0$,
$c_{2,3,4}\sim\frac{m}{\Lambda}c_1$ are small compared to
$c_1$.  Both lattice artifacts and the contributions of
$O(m^2)$ terms can be ignored. In this case $c_1 \propto m$,
a result we use several times below. In fact, the proportionality
constant ($\alpha_1 \Lambda^2$) is very likely to be positive,\footnote{%
In the unquenched theory $\alpha_1$ (which is related to the
Gasser-Leutwyler parameter $B_0$) is known to be positive, and we do not
expect quenching to change the sign of this parameter. Numerical evidence
supports this expectation.}
and we phrase subsequent discussions as though this is true,
although our final conclusions would not change if $\alpha_1$
were negative.
\item[2.]  The quark mass is $O(a)$ itself, \ie\
$am\sim(a\Lambda)^2$. In this case, $c_{2,3,4}$ are still small compared
to $c_1$, but the $O(a)$ term in $c_1$ cannot be ignored.
The critical value of $m$ is shifted by an amount 
$\sim a\Lambda^2$, and one may define $m'$ as a subtracted
quark mass such that the pion mass vanishes at $m'=0$.
The two terms in $c_1$ are both of the same order.
\item[3.] The subtracted quark mass is of order $am'\sim(a\Lambda)^3$.
In this case, we have that $c_1\sim m'\Lambda^3$ and
$c_{2,3,4}\sim a^2\Lambda^6\sim c_1$.
In other words, for this case the higher-order terms in
$V(\Sigma)$ compete with the lower-order term, and we need
to consider both to determine the vacuum structure of the
theory.  As already observed in Ref.~\cite{shsi}, this leads
to the prediction that the width of a potential Aoki phase
is of order $a^3$ (for small $a$).  In this regime, higher-order
terms beyond those shown in Eq.~(\ref{POTENTIAL}) are of higher
order in $a$, and therefore need not be included in the analysis
\cite{shsi}.
\end{itemize}

In summary, the quenched approximation has introduced several new features
to the potential: traces have become supertraces;
there are three $O(p^4)$ terms
in the potential rather than one, due to the different group structure;
there are extra terms involving $\Phi_0$;
and, finally, the requirement to do an axial rotation has led to the
leading mass-like term have the form $\Sigma - \Sigma^{-1}$ rather
than $\Sigma+\Sigma^{-1}$. We now turn to the question of how these
changes influence the phase diagram.

\section{Phase Diagram}
\label{sec:phasediagram}

We now address the central question of this paper,
whether there can be an Aoki phase in the quenched theory,
analogous to that predicted for the unquenched two-flavor theory.
This might seem to be just a matter of minimizing the potential we
have constructed, but there are a number of subtleties
in the ghost sector that complicate the analysis.
In fact, we will need certain additional properties of the
quenched theory, which we collect before proceeding to the full analysis.

\subsection{Additional Properties of the Quenched Theory}

To study the spontaneous breaking 
of parity and the flavor symmetry group $\cH$ of Eq.~(\ref{FULLVECTOR}), 
we will only allow source terms which respect the graded symmetry between
a quark and its corresponding ghost quark.  The reason is, of course, that
ghost quarks are not present in numerical simulations, but instead the
quark determinant is simply omitted.  The ghost-quark is just a 
field-theoretical trick to describe this procedure in terms of a 
path integral, and our choice of sources should be restricted 
accordingly.  Furthermore,
it is sufficient to use source terms which are diagonal in flavor.
This is because we can choose any flavor-breaking condensate
to point in the $\tau_3$ direction.
The fact that the source terms are flavor-diagonal then implies
that quarks of different
flavors are not coupled (either by the action or the sources).
This in turn means that quenched correlation functions involving
only one flavor of quark depend only on the mass of that quark,
and not on those of other quarks, or even their presence or absence.
It also follows that correlation functions involving ghost quarks are
identical, up to a possible overall sign, to those for the corresponding
quark, since they are composed of the same propagators.
These results hold for any lattice spacing and thus also in the continuum 
limit.

A consequence of these observations is that the quenched condensate
for each flavor is independent of the number of flavors, and that 
any ghost condensate is equal to that of the corresponding quark.
This holds not only for the bare lattice condensates,
but also for the physical condensates. The latter are obtained by
subtracting divergent contributions from the bare condensate,
and then multiplying by a matching factor. These steps maybe difficult
to implement in practice, but what matters to us here is that they
can be done in principle, and that the subtractions and matching factors
do not depend on the number or properties of flavors other than
that in the condensate itself.  It follows in particular that any
spontaneous symmetry breaking pattern does not break symmetries
in $\cH$ such that any ghost-quark condensate would be different from
the corresponding quark condensate.

Our next observation concerns the quenched condensate in the continuum
limit. In this limit, one can ignore the $b_1$ term in
Eq.~(\ref{QLEFFQUARK}), and the quark condensate is an odd function
of the quark mass, configuration by configuration:
\begin{equation}
\tr[1/(D+i\gamma_5 m)] = \tr[\gamma_5^2/(D+i\gamma_5 m)]
=-\tr[1/(D-i\gamma_5 m)]
\,.
\label{SIGNFLIP}
\end{equation}
If the condensate has a non-vanishing
non-perturbative contribution, as we are assuming, then this
too must be odd in $m$. The same holds for the corresponding
ghost-quark condensate. 
Such a behavior is indeed seen in
numerical results for the quenched condensate with staggered 
and overlap fermions, which is odd under $m\to -m$.
At the level of the effective chiral
lagrangian, working in the continuum limit means that
$c_1 \propto m$ and $c_{2,3,4}\propto m^2$ in
Eq.~(\ref{POTENTIAL}), 
and the result (\ref{SIGNFLIP}) implies
 that $\Sigma_{qq}$
(the component of the condensate for quark $q$),
and the corresponding ghost condensate $\Sigma_{gg}$,
should, in the continuum limit,
both change sign when $c_1$ does.
This result also follows from symmetries in the following
somewhat indirect way. First, for the condensate in the
physical sector, $\Sigma_{qq}$, this  can be derived using the
appropriate axial symmetry in order to flip the sign of $m$.
Then, because of the arguments given above, the corresponding
ghost condensate, $\Sigma_{gg}$, has to follow suit.
We note that there is no
``direct" way of changing the sign of $m$ in the ghost sector,
as follows from Eq.~(\ref{GHSPURION}).
We note also that this result does not hold in the unquenched
theory because the measure depends on the quark mass.

In light of the previous discussion,
one might wonder why we need to consider more than one flavor
in the quenched theory. The reason we need two flavors is to
allow the calculation of flavor non-singlet pion propagators.
These have only quark-connected contributions and are the
quantities usually calculated in simulations. It is the masses
of these pions which are observed to extrapolate to zero at
non-zero lattice spacing --- the phenomenon which the Aoki phase
was introduced to explain. With one flavor alone one can only
consider flavor singlet pions, which have quark-disconnected
contributions as well.

We also need to recall some properties of the quenched theory 
in the limit that $N_c$, the number of colors, becomes large.%
\footnote{%
One subtlety of this limit is that we must take $N_c\to\infty$ before
$m\to0$, so that the
quenched artifacts proportional to $\ln(m)/N_c$ 
vanish. Since we are implicitly working 
at small but non-vanishing quark masses, so as to avoid these artifacts,
we are able to take the large-$N_c$ limit without problems.}
In the continuum effective lagrangian, Eq.~(\ref{LEFFANOM}), 
standard arguments show that
the constants $f^2$ and $v$ are proportional to $N_c$, while
$c_0$ and $\alpha$ are $O(1)$.
The latter two are suppressed because they multiply terms
corresponding to disconnected diagrams at the quark level.
Similarly, in the potential for non-zero lattice spacing, 
Eq.~(\ref{POTENTIAL}),
the coefficients of the
single-supertrace terms, $c_1$ (which is proportional to $vm$) and
$c_4$, are $O(N_c)$, while the coefficients of terms with two supertraces,
$c_0$, $c_2$ and $c_3$, are $O(1)$.
Thus in the large-$N_c$ limit we can set $c_0$, $c_2$ and $c_3$ to zero.

The advantage of this limit is, of course, 
that the quark sector becomes physical,
since the quark determinant is suppressed by $1/N_c$.
In other words, there is no distinction between quenched
and unquenched theories in this limit.
Thus the results of our analysis of the
quenched theory must, in the quark sector,
coincide with those of the unquenched two-flavor
theory in this limit. The analysis of the Aoki phase
in the unquenched theory in the large-$N_c$ limit is
a simple generalization of that at finite $N_c$~\cite{shsi}.
In particular, the competition between two contributions which
leads to the non-trivial phase structure remains in the
large-$N_c$ limit (since these are the $c_1$ and $c_4$ terms).
The only significant change is that the flavor group
$SU(2)$ is enlarged to $U(2)$. This larger symmetry group
may then be broken down to $U(1)\times U(1)$ (as opposed to a single
$U(1)$) in the Aoki phase, so there are still two exact Goldstone
bosons.  Further details will be discussed below.

The particular utility of the large-$N_c$ limit is that it allows
us to check our  method of analysis of the ghost sector.
We have argued above that, with sources appropriately chosen, 
the quark and ghost condensates must be the same.
Since the analysis in the quark sector becomes unquenched
as $N_c\to\infty$, this means that we know the result which
we must obtain in the ghost sector in this limit.

\subsection{Phase Diagram as $\mmbf{N_c\to\infty}$}

We start from the effective potential $V(\Sigma)$ of Eq.~(\ref{POTENTIAL})
for two flavors, $u$ and $d$,
and substitute the form for $\Phi$, Eq.~(\ref{PHIPAR}). 
In fact, we can simplify this form as follows: first, by
setting $\chi=\chibar=0$, since these are the
solutions of the classical equations of motion for these fields;
second, by setting $\phi_2=0$, as discussed in the previous section;
and, third, by keeping only flavor-diagonal components of the fields since
we force any flavor-breaking to lie in the $\tau_3$ direction.
Thus we can investigate the vacuum structure using an expectation value
\begin{equation}
\label{sigpar}
\langle \Sigma  \rangle \equiv 
\Sigma = \mathrm{diag}(e^{i\phi_u}, e^{i\phi_d}, e^{\phibar_u}, e^{\phibar_d})
\,,
\end{equation}
where $-\pi<\phi_{u,d}\le \pi$ are real phases,
while $\phibar_{u,d}$ are real variables to be integrated along the 
entire real axis.

In this subsection we consider the large-$N_c$ limit, and thus set 
$c_0=c_2=c_3=0$.
The potential is then
\ba
V &=& 2 \sum_{j=u,d} \Biggl[ {\rm sgn}(\epsilon_j)
c_1 \sin(\phi_j) -|\epsilon_j| \cos(\phi_j)
+ c_4 \cos(2 \phi_j)  \label{VLARGENC} \\
&& \hspace{1cm}  + i\; {\rm sgn}(\epsilon_j) c_1 \sinh(\phibar_j) 
+|\epsilon_j| \cosh(\phibar_j)
- c_4 \cosh(2 \phibar_j) \Biggr]
\,. \nonumber
\ea
Here we have reintroduced the dependence on the $\epsilon_j$, which follows from
the fact that the lattice mass and Wilson terms in Eq.~(\ref{LLATT}) are
proportional to ${\rm sgn}(\epsilon_j)$. The $c_4$ term does not depend
on the sign of $\epsilon_j$ since it is quadratic in symmetry breaking. 
The factorization into four parts, dependent respectively
on $\phi_u$, $\phi_d$, $\phibar_u$ and $\phibar_d$, is the result
of taking $N_c\to\infty$. This is how, in this context, the quark
and ghost sectors decouple.

In the quark sectors the analysis is essentially a 
recapitulation of that of Ref.~\cite{shsi} 
for the unquenched two-flavor theory,
except that here it is done for each flavor separately.
The notation is also different: our $c_4$ was called $-c_2/4$
in Ref.~\cite{shsi}, and here we have done an axial transformation
which complicates the interpretation of the condensate.
What remains unchanged is the region of parameters of interest:
$c_1$ is proportional to the $O(a)$ shifted quark mass $m'$
while $c_4\sim a^2$ is a constant. A useful ratio is
$\eta = c_1/4 c_4$. We want to determine
what happens to the condensate as $\eta$ ranges from
values smaller than $-1$ to larger than $1$, in the limit
$\epsilon\to 0$.
We note that the term proportional to $|\epsilon_j|$ can be
dropped for most of the analysis, since we take
$\epsilon_j\to0$ at the end, except where it is needed
to distinguish between otherwise degenerate minima.

It is useful, first, to
consider the continuum theory in which $c_4=0$,
so that the potential in each quark sector is
$V_q =2\, {\rm sgn}(\epsilon_q) c_1 \sin(\phi_q)$,
where we now use an index $q$ instead of $j$ to
indicate that we are considering the quark sector
of a given flavor~$q$. 
For positive $c_1$ (corresponding to positive
quark mass in the original basis)
the minimum is at $\phi_q=-{\rm sgn}(\epsilon_q)\pi/2$, so that
$\Sigma_{qq}=e^{i\phi_q}=-{\rm sgn}(\epsilon_q)i$. 
For negative $c_1$, the
minimum switches to $\phi_q={\rm sgn}(\epsilon_q)\pi/2$, so that
$\Sigma_{qq}={\rm sgn}(\epsilon_q) i$. In short, the minimum
occurs at $\Sigma_{qq}=-{\rm sgn}(\epsilon_q \, c_1)i$. 
This reproduces the standard discontinuous
dependence of the direction of the
condensate as a function of the direction of the mass and source terms.

The unusual factor of $-i$ in the condensate
is a reflection of the axial transformation Eq.~(\ref{TRANSF})
is needed to define the quenched theory.
If we rotate back to the usual continuum basis, then the
condensate becomes $\Sigma_{qq}={\rm sgn}(c_1)$.
In other words the non-standard values of $\phi_q$ 
``undo" the axial transformation. To see this explicitly, we can
write the potential in terms of shifted fields
$\phi_q =-{\rm sgn}(\epsilon_q \, c_1)\, \pi/2+\phi'$: 
\be
\label{QPOT}
V_q = -2 |c_1|\cos{\phi'} - 2 c_4 \cos{2\phi'}
= -2(|c_1|\! + \!c_4) + (|c_1|\! +\! 4 c_4) \phi'^2 
+ O(\phi'^4)
\,,
\ee
where we have restored the $c_4$ term in anticipation of the 
discussion below. 
This is the potential we would have obtained directly
if we had not had to use the trick with the
axial transformation of Eq.~(\ref{TRANSF}), and instead expanded about
$\Sigma_{qq}=\pm 1$.

Now we consider non-vanishing $c_4$. If $c_4 > 0$, the analysis 
with $c_4=0$ goes through unchanged: the $c_4 \cos(2\phi_q)$
term has equal minima at $\phi_q=\mp\pi/2$, so it is the
sign of $\epsilon_q c_1$ which determines the direction
of the condensate. There is thus
a first order transition when $c_1$ changes sign.

If $c_4<0$, however, it contributes negative curvature
to $V_q$, which can destabilize the minima. 
Looking at the quadratic terms in Eq.~(\ref{QPOT}),
we see that this happens when $4 c_4 < -|c_1|$,
\ie\ $|\eta|<1$.
The minimum of the potential shifts to 
$\phi_q= {\rm sgn}(\epsilon_q)\arcsin(\eta)$, signaling that
we are in an Aoki phase.
There are two choices of branch for the inverse sine,
one interpolating between $-\pi/2$ and $+\pi/2$, the other between
$-\pi/2$ and $-3\pi/2$. 
In both cases the condensate ``swings" from $-i$ to $+i$ around
the unit circle in the complex plane, when $\epsilon_q c_1$ changes from
positive to negative values.
The choice of branch decides which way the condensate swings,
\ie\ whether it passes through $+1$ or $-1$ when $\eta=c_1=0$.
Which choice is appropriate is determined by the
$|\epsilon_j|$ term in Eq.~(\ref{VLARGENC}).  
This term is negative if $-\pi/2 <\phi_q < \pi/2$ and positive
if $ -3\pi/2 < \phi_q < -\pi/2$, and thus selects the branch in which
$\Sigma_{qq}$ swings through $+1$. This is true irrespective of
the sign of $\epsilon_j$.

The symmetry that is broken by the condensate depends on the
details of the theory. If we were considering
one flavor alone, then, undoing the axial transformation (\ref{TRANSF}),
we would find that $\langle \qbar \, i \gamma_5 q \rangle \propto 
{\rm sgn}(\epsilon_q)\sqrt{1-\eta^2 }\ne 0$ for $|\eta| < 1$.
Thus parity is broken in the Aoki phase. The sign of the condensate
is determined by the sign of $\epsilon_q \equiv \epsilon_j $ in Eq.~(\ref{POMASS}).
There are no exact
Goldstone bosons in this phase as no continuous lattice symmetry
is broken.

Next, we consider the two-flavor theory, with degenerate masses.
The pattern of symmetry breaking 
depends on the signs we choose for $\epsilon_u$ and $\epsilon_d$ in
Eq.~(\ref{POMASS}).
If we choose $\epsilon_u$ and $\epsilon_d$ both of the same sign, so that
Eq.~(\ref{POMASS}) represents 
a flavor singlet source proportional to $\sum_q \qbar \,i \gamma_5 q$,
then the condensates will satisfy $\langle \bar u \,i \gamma_5 u 
\rangle = \langle \bar d \, i \gamma_5 d \rangle \ne0$,
and parity, but not flavor, will be broken 
(so again there are no Goldstone bosons).
We checked explicitly, from Eq.~(\ref{POTENTIAL}) with $c_0=c_2=c_3=0$, that in this
case indeed there are no Goldstone bosons (all pions have a mass of order $a$).
If, on the other hand, we use a flavor-breaking source term proportional to
$\bar u \, i \gamma_5 u - \bar d \, i \gamma_5 d$ (\ie\ $\epsilon_d=-\epsilon_u$), 
the mass term in the lattice quark lagrangian will be proportional
to $\tau_3$ (\cf\ Eq.~(\ref{QTWOFLAVOR})), which translates into the 
$c_1$ terms in Eq.~(\ref{VLARGENC}) for the up and down sectors having 
opposite signs. The condensates will then satisfy
$\langle \bar u \, i\gamma_5 u \rangle = -\langle \bar d \, i \gamma_5 d 
\rangle \ne0$, and flavor and parity will be
broken. This is the most familiar example of the Aoki phase, which has two
Goldstone bosons $\pi^\pm$, and which carries over to finite $N_c$
in the unquenched theory (see Sect.~5.3 below).
In this case, the calculation of the meson spectrum from the chiral
lagrangian is virtually identical to that in Ref.~\cite{shsi}, and
confirms the existence of two massless Goldstone bosons.
We stress, however, that, as far as the condensate is concerned,
these theories at infinite $N_c$ differ
only kinematically --- the underlying phenomenon is identical in all cases.
\bigskip

We now turn to the ghost sector, in which the potential is
\begin{equation}
V_g = 2 \left[i\, c_1^\prime \sinh(\phibar) 
+ |\epsilon| \cosh(\phibar) - c_4 \cosh(2 \phibar) \right] \,.
\label{GHPOT}
\end{equation}
We have dropped the index $j$ because, in the remainder of this
section, we are only interested in demonstrating that, for each
flavor, the ghost sector follows the quark sector. We have introduced
the useful variable\footnote{%
We could have performed the analysis in the quark 
sector in terms of $c_1^\prime$ as well; however, there we wanted 
to emphasize the explicit flavor dependence of the condensate in 
the Aoki phase through the signs of the $\epsilon_q$.
}
$c_1^\prime \equiv {\rm sgn}(\epsilon_j) c_1$.
As in the quark sector, we can ignore the $|\epsilon|$ term except
when it breaks degeneracies.

The potential (\ref{GHPOT}) is complex, and the usual approach of minimizing
it to find the vacuum structure breaks down.  However, what one
is really doing if one minimizes the potential in the context
of euclidean field theory is to calculate the leading term 
in a saddle-point expansion of the path integral.  Stating the
problem this way, it is clear that this can also be done
when the integrand is complex instead of real.  The prescription
is to treat $\phibar$ as complex, and deform the contour
of integration so as to pass through a saddle-point.
The vacuum expectation value of
the field is, at leading order, the value at the position of
the saddle.
If there are multiple saddles, the appropriate one to use is
determined by the boundary conditions on the contours, and
other considerations, as will be discussed below.

Another issue we must face is the convergence of integrals
as $\phibar\to \pm\infty$. For instance, if $c_4>0$, 
the last term in $V_g$ is unbounded
from below in these limits. 
In fact, we only know the form of the potential for small fields,
$|\phibar| \lesssim 1$. The ratio of successive terms in the chiral
expansion grows exponentially when $|\phibar|$ exceeds unity
(as can be seen from the relative magnitude of the $c_4$ and $c_1$ terms)
rapidly overcoming any suppression in their coefficients.
Thus, once $|\phibar|\sim 1$ the chiral
expansion breaks down, and we do not know its behavior at 
large fields.
In light of this, we simply assume that we can treat the integral over 
$\phibar$
as convergent at infinity along the real axis, and search for saddles in
the region $|\phibar|\leqx 1$ through
which the contour can be deformed in such a way that it can be evaluated using
steepest descent.
Note that the same issues do not arise in the $\phi_q$ integrals, because there
is no relative exponential growth in the terms, and because the integral runs
over only a short segment (of length $2\pi$) of the real axis.

As for the quark sector, we warm up by considering the phase structure
for $c_4=0$, so that  $V_g=2i c_1^\prime \sinh(\phibar)$.
The saddle point equation is 
\begin{equation}
\cosh(\phibar)=\cosh(x)\cos(y) + i \sinh(x)\sin(y) = 0
\,,
\end{equation}
where we have written $\phibar=x + i y$, with $x$ and $y$
real.
The solutions to this equation are a periodically repeating sequence 
along the imaginary axis, with the
two nearest to the origin being at
$\phibar_A=-i\pi/2$ and $\phibar_B=+i\pi/2$. These correspond to the condensate 
taking the values $\Sigma_{gg}=\exp(\phibar_{A,B})=\mp i$, \ie\ the same values as arose
in the quark sector. The other solutions lead to the same two values of the 
condensate and are thus not physically distinct.

Expanding about the saddles, we find
\begin{equation}
V_g(\phibar=\mp i \frac{\pi}{2} + \phibar')
=
\pm 2 c_1^\prime \cosh(\phibar') 
=
\pm 2 c_1^\prime \left[1 + \frac{\phibar'^2}{2}\right]\,.
\end{equation}
The criteria we use to choose the saddle are based on the direction in
which $\Re V_g$ rises most steeply (which is the direction the
contour should follow to have the steepest descent of $|\exp(-V_g)|$
away from the saddle), as well as the value of the potential at the saddle. 
The contour should pass through the saddle and then be able to join
the real axis for large fields, without passing near other saddles.
If there is more than one saddle point with an appropriate contour,
we want to {\em maximize} the value of $\Re V_g$, so that the contribution
to the saddle-point
integral is minimized, and thus a better approximation to the (absolute)
value of the full integral is obtained.%
\footnote{For a lucid description of the saddle-point method, see
\eg\ Ref.~\cite{bruijn}.}
All criteria agree in this case: if $c_1^\prime>0$ then 
$V_g$ is larger at $\phibar_A$, and has steepest descent in the direction
of $\phibar'$ real, so that the contour can be easily deformed so as to
end up on the real axis for large $\phibar'$.
The other saddle has its direction of steepest descent along the imaginary 
axis,
with the contour heading directly for saddle $\phibar_A$ and its periodic
reflection.
If $c_1^\prime< 0$, the role of the two saddles is reversed, and we must choose
$\phibar_B$.

In this way, the ghost condensate is predicted to flip from $-i$ to
$+i$ as $c_1^\prime$ passes from positive to negative values.
This is exactly the same dependence as for the quark condensate, $\Sigma_{qq}$,
and is in agreement with the general arguments given in Sect.~5.1.
That this check works out gives us confidence in our method.
We note for future reference
that the total potential, $V_q+V_g$, vanishes at the saddles
that have been chosen.

Now we add back in a non-zero $c_4$, in which case the saddle point equation 
becomes
\begin{equation}
i c_1^\prime \cosh(\phibar) - 2 c_4 \sinh(2 \phibar) = 0
\,.
\end{equation}
The solutions are $\phibar_{A,B}$, defined above, and an additional
saddle or saddles satisfying
\begin{equation}
\phibar_C: \qquad \sinh(\phibar_C) = i c_1^\prime/(4 c_4) \equiv i 
\eta^\prime \,.
\end{equation}
The solutions of this equation (apart from periodic repetitions in the
imaginary direction) are
\begin{eqnarray}
\eta^\prime < -1:&&
\phibar_{C\pm}= -i\pi/2 \pm \arcosh(-\eta^\prime) =\phibar_A\pm \arcosh(-\eta^\prime) 
\label{saddles}\\
-1 \le \eta^\prime \le +1:&&
\phibar_C = i \arcsin\eta^\prime \nonumber\\
\eta^\prime > 1:&&
\phibar_{C\pm}= i\pi/2 \pm \arcosh\eta^\prime =\phibar_B\pm \arcosh\eta^\prime 
\,. \nonumber
\end{eqnarray}
Expanding $V_g$ around these saddles we find
\ba
V_g(\phibar_A+\phibar') &=& 2 (c_1^\prime+c_4) + 4 c_4 (\eta^\prime+1) \phibar'^{\,2} 
+ O(\phibar'^{\,3}) \,,\label{Vsaddle}\\
V_g(\phibar_B+\phibar') 
&=& 2 (-c_1^\prime+c_4) + 4 c_4 (-\eta^\prime+1) \phibar'^{\,2} 
+ O(\phibar'^{\,3}) \,,\nonumber\\
V_g(\phibar_C+\phibar')
&=& - 2 c_4 (1 + 2 \eta^{\prime\,2}) 
+ 4 c_4 (\eta^{\prime\,2} - 1) \phibar'^{\,2} 
+ O(\phibar'^{\,3}) \,. \nonumber
\ea
As above, our approach will be to find the saddle for which $V_g$ increases
for real displacements 
(which we will refer to as ``positive curvature"), since the original
integral over $\phibar$ is along the real axis,
and which has the largest value of the potential at the saddle.
We run through the choices of parameters in turn. 

\bigskip
We begin by considering $c_4>0$.

\medskip\noindent
{$\mathbf{c_1^\prime> 4c_4 >0}$, or $\eta^\prime>1$:}  
Saddles $A$ and $C\pm$ (arrayed either side of $B$)
have positive curvature, while
$B$ has negative curvature.
There are two choices of contour: that passing through $A$ alone,
and that running through  $\phibar_{C-}$,
$\phibar_B$ and then $\phibar_{C+}$. We 
choose the former as the potential is higher at $A$ than at $C\pm$.
Furthermore, saddles $C\pm$ are outside the range
of applicability of our truncated effective potential
except for $\eta^\prime\approx 1$.
Thus we find the same saddle as for $c_4=0$.

\bigskip\noindent
{$\mathbf{4 c_4> c_1^\prime >0}$, or $0<\eta^\prime<1$:}
Both saddles A and B have positive curvature, while $C$ lies
on the imaginary axis between A and B and has negative curvature.
Thus a contour passing through either A or B is possible.
We choose that through A since it has the higher potential.

\medskip\noindent
{$\mathbf{4 c_4> 0 > c_1^\prime > -|4 c_4|}$, or $-1<\eta^\prime<0$:}
The saddles are similar to the previous case, except that
saddle B now has the highest potential, and so we pick the
contour passing through B rather than A. 
In other words, when $c_1^\prime$ passes through zero the saddle
switches from A to B, just as in the case $c_4=0$.

\medskip\noindent
{$\mathbf{4 c_4> 0 > -4c_4 > c_1^\prime}$, or $\eta^\prime<-1$:}
The situation is the reflection in the real axis of that
for $c_1^\prime>4 c_4 >0$. Saddles $B$ and $C\pm$ have positive
curvature, and we choose the former since it has the higher
potential.

\medskip\noindent
{\bf Summary for $\mathbf{c_4>0}$:}
The ghost condensate is unchanged from the analysis
when $c_4=0$, and satisfies $\phibar = i \phi$
for all $c_1^\prime$. Thus $\Sigma_{qq}=\Sigma_{gg}$, as expected.
In both quark and ghost sectors there is a first-order
phase transition as $c_1^\prime$ passes through zero, just
as in the continuum limit. There is no Aoki phase in
either sector. The spectrum of fluctuations about
both quark and ghost vacua is the same for all $c_1^\prime$.

\bigskip
We now turn to the case $c_4<0$.

\medskip\noindent
{$\mathbf{c_1^\prime> 4|c_4| >0 > 4 c_4}$, or $\eta^\prime<-1$:}
Changing the sign of $c_4$ changes the sign of the curvature about all saddles.
Thus, in this parameter range, only
saddle $A$ has positive curvature. The contour
must pass through $A$, and then move back to the real axis.
It can do so by passing through $C\pm$ and then moving off
in an imaginary direction, although this is not necessary.
In all cases the integral is dominated by the value at saddle $A$.

\medskip\noindent
{$\mathbf{ 4|c_4|> |c_1^\prime|\,;\ \  0 > c_4 }$, or $|\eta^\prime|<1$:}
The contour must pass through saddle $C$, since this is the only
saddle with positive curvature. 
This gives an expectation value $\phibar= i\arcsin\eta^\prime
= i\;{\rm sgn}(\epsilon_j)\arcsin\eta$,
which is consistent with the equality $\phibar=i\phi$,
since we found above that $\phi_q = {\rm sgn}(\epsilon_j)\arcsin\eta$.
As for the quark sector, however,  there is a choice of branch of the
inverse sine function. The appropriate choice is determined by
the $2 |\epsilon_j|\cosh\phibar|$ term in $V_g$: one should pick the branch
which {\em maximizes} this term. It is easy to see that the resulting
branch is that which satisfies the equality  $\phibar=i\phi$ 
 for either sign of $\epsilon_j$, \ie\ that the ghost condensate
is always equal to the quark condensate.

\medskip\noindent
{$\mathbf{0 > 4c_4 > c_1^\prime}$, or $\eta^\prime>1$:}
The contour passes through $B$,
which is the only saddle with positive curvature.

\medskip\noindent
{\bf Summary for $\mathbf{c_4<0}$:}
As for $c_4>0$, these results are consistent with
those from the quark sector and give
$\Sigma_{gg}=\Sigma_{qq}$. In particular,
we find an Aoki phase in both quark and ghost sectors.

\bigskip\bigskip
In summary, we see how the saddle-point analysis effectively
restores the quark-ghost symmetry which had been broken by
the need for convergence of the ghost integral.

\subsection{Phase Diagram for Finite $\mmbf{N_c}$}

We now extend our analysis of the quenched phase
diagram to finite $N_c$.  Finite-$N_c$ corrections will 
change the values of $c_1$ and $c_4$, but, more importantly, turn
on $c_0$, $c_2$ and $c_3$.  The latter terms couple the quark and
ghost sectors, as well as the different flavors.
To make progress here we will need to use
the assumption, discussed in Sect.~5.1,
that the condensates in the quark and ghost sectors
have to be equal, as well as other general considerations.

We note first that it is straightforward to show that the solutions we found in
Sect.~5.2 are still solutions of the full saddle-point equations 
which follow from Eqs. (\ref{POTENTIAL}) and (\ref{sigpar}).  
The full saddle-point equations in the finite-$N_c$ case are
transcendental equations, due to the $c_0$ term in the effective
potential $V(\Sigma)$, and we do not know whether any solutions
to these equations other than those discussed in the previous section
exist.  
However, in the previous section we found that the ghost and quark 
condensates were always equal in the $N_c \to \infty$ limit, and,
in accordance with the argument at the quark level in Sect.~5.1, 
we will insist that this continues to hold for finite $N_c$. This 
means that we require $\phibar_q=i\phi_q$ 
for each flavor $q$. Remarkably, upon imposing this constraint
we find that the terms proportional to $c_0$, $c_2$ and $c_3$
drop out of the saddle-point equations, and the solutions are 
those found in Sect.~5.2. Note that a similar argument applies 
to the terms of order $\Phi_0^2$ and beyond\footnote{These terms 
all vanish in the large-$N_c$ limit.}
in the potentials $V_{0,2}(\Phi_0)$ in Eq.~(\ref{LEFFANOM}).

We conclude that reducing $N_c$ from infinity does not change the saddles
we need to consider. This is the good news. The bad news is that,
since we are now considering the coupled $\phi_q$, $\hat\phi_q$ system, we
must use the full potential when distinguishing between the saddles,
and, as noted above, this potential vanishes for all of them.

To make progress, we consider in general the role of the $c_0 \Phi_0^2$ term
in the quenched chiral potential (\ref{POTENTIAL}).
A signature result of quenched chiral perturbation theory is that,
at tree level,
this term does not shift the position of the pole in the flavor-singlet
propagator but rather gives rise to a double-pole contribution at the
same position.  This double-pole term represents the presence of 
quark-disconnected contributions (the so-called ``double hairpins")
in this channel \cite{bg,sharpe}.%
\footnote{We expect this to be true inside the Aoki phase as well.}
Furthermore, the $c_0$ term does not affect the value of the condensate
at tree level, but rather only at one-loop through long-distance effects.
This is very different from the effect of such a term in the unquenched
theory. There it shifts the position of the $\eta'$ pole, and changes
the value of the condensate at leading order (since $\eta'$ loop effects 
are short distance they lead to an $O(1)$ correction
to the condensate proportional to $v\times M_{\eta'}^2/(4\pi f_\pi)^2$). 

We conjecture that these perturbative results
hold also non-perturbatively, and that the $c_0$ term does not 
influence the value of the condensate.
A similar discussion applies to $c_2$ and $c_3$ terms in the
potential, which both have two supertraces, and at leading order in a field
expansion are proportional to $\Phi_0^2$.
Given this conjecture, the analysis for finite $N_c$ collapses to that
for infinite $N_c$, and we conclude that the possible
phase structures in the quenched theory at finite  $N_c$
are the same as those in the unquenched theory at infinite $N_c$.

We illustrate the argument for the irrelevance of
$c_0$, $c_2$ and $c_3$ by studying the small fluctuations
around one of the saddles. 
For simplicity, assume
that $c_1^\prime>0$ and that $c_1^\prime+4c_4>0$ as well, so that 
we are in the phase with $i\phi_q=\phibar_q=-i\pi/2$ (this is 
saddle point $A$ in the ghost sector).  Expanding the effective
potential around this saddle to quadratic order gives 
\begin{equation}
\label{VQUADR}
V(\Sigma)=\frac{1}{2}\pmatrix{\phi_q,&\phibar_q}
\Big[2(c_1^\prime+4c_4){\bf I}+(c_0+4c_2+2c_3){\bf X}\Big]
\pmatrix{\phi_q\cr\phibar_q}\,,
\end{equation}
in which ${\bf I}$ is the unit matrix and ${\bf X}$ is a matrix filled
with $1$'s in the quark sector (upper left-hand block), $i$'s
in the mixed quark-ghost sector, and $-1$'s in the ghost sector
(lower right-hand block). Note that the coefficient of ${\bf I}$
reproduces the coefficients of the quadratic terms in the first 
line of Eq.~(\ref{Vsaddle}), as well as Eq.~(\ref{QPOT}) for the 
case at hand. 

First, ignoring  ${\bf X}$, we see that the potential is stable 
around our solution if \hbox{$c_1^\prime+4c_4>0$}, with an instability
developing when $c_1^\prime+4c_4<0$, signaling spontaneous symmetry
breaking.  This confirms what we found in Sect.~5.2.  As long
as $c_1^\prime+4c_4>0$, $\phi_q$ and $\phibar_q$ mesons propagate
with masses proportional to $c_1^\prime+4c_4$.  

Now if we include ${\bf X}$, the (zero-momentum)
propagator can be determined from the inverse of the matrix in
Eq.~(\ref{VQUADR}).  Because ${\bf X}^2=0$, this
inverse is equal to
\begin{equation}
\label{INVERSE}
\frac{1}{2(c_1^\prime+4c_4)}\,{\bf I}-\frac{c_0+4c_2+2c_3}
{4(c_1^\prime+4c_4)^2}
\,{\bf X}\,.
\end{equation}
This shows that the parameters $c_{0,2,3}$ do not affect the
meson masses, but instead determine the residue of the quenched
double pole, which originates in the so-called ``double-hairpin"
diagram.  While this double
pole exhibits a sickness of the quenched theory, it does not
affect the meson masses.  Spontaneous symmetry breaking only
occurs when the squares of these masses turn negative.

This brings us to our final point of this section, 
which is the effect of the flavor-singlet pseudoscalar
in the unquenched theory.  For simplicity we discuss only
the two-flavor theory. In the unquenched
theory, the flavor-singlet mass term is $(1/2)c_0\Phi_0^2
=(1/2)c_0(\phi_u+\phi_d)^2$.  In the limit of vanishing
$\epsilon_{u,d}$ this term raises the energy of the solution
for which $\phi_u=\phi_d$.  Thus, in the unquenched theory,
one finds that $\langle \bar{u}\, i \gamma_5u \rangle 
=-\langle \bar{d}\, i \gamma_5d \rangle \ne 0$ inside the
Aoki phase, and flavor is always broken along with parity.
As we have seen,
this mechanism for picking the vacuum is not present in
the quenched theory.  In the quenched theory, for both
infinite and finite $N_c$, the pattern of symmetry breaking 
depends on the relative sign of $\epsilon_u$ and $\epsilon_d$.

\section{Conclusion}
\label{sec:concl}

The aim of this paper is to see whether the method
of Ref.~\cite{shsi}, 
based on the effective theory describing the
Goldstone-boson physics of full QCD with two flavors of Wilson
fermions, can be extended to the quenched theory.
Our conclusion is that it can, although to push through
the analysis we have had to make
a number of assumptions not needed in the unquenched theory.
We find the same two possibilities for the phase structure
as in the unquenched case: depending on the sign
of a parameter of the effective theory ($c_4$), there is either
an Aoki phase or a first-order phase transition. 

The nature of the quenched Aoki phase, if there is one,
can however differ from that in the unquenched theory.
In particular, the form of the condensate in the two-flavor quenched
theory depends on the source employed to probe spontaneous
symmetry breaking.  One can have a phase in which both
parity and flavor symmetry are broken, as in Aoki's original
scenario, but one can also have only breaking of parity,
with no breaking of flavor, and thus no Goldstone bosons.
Since in the quenched theory quark correlators are only
probing the theory, while not being part of the dynamics
(there are no sea quarks),
this is a kinematical effect.  For more detail, we refer
to Sect.~5.2.  Unquenched QCD at $N_c=\infty$ shares this
property, but there the degeneracy between the two possibilities
is lifted at finite $N_c$, in accordance with Ref.~\cite{shsi}.

The extension of the method of Ref.~\cite{shsi} to the
quenched theory turned out to be non-trivial.
The formal definition of quenched QCD of
Ref.~\cite{bg} does not lead to a convergent path integral
in the ghost sector.  While expanding the lagrangian of
Ref.~\cite{bg} around $\Sigma=1$ leads to the correct
version of quenched ChPT, it breaks down non-perturbatively.
We gave a non-perturbatively valid path-integral definition
of quenched QCD, and analyzed its symmetries.  This has been
discussed before in the continuum in Ref.~\cite{rm},
but here we extended this to lattice QCD with Wilson fermions,
thus providing a fully-regulated, convergent path-integral
definition of quenched QCD.
We then used it to construct an effective potential to
next-to-leading order, including terms up to order $a^2$.
Employing this effective action, we argued that the
analysis of the phase structure of unquenched QCD with
two Wilson fermions of Ref.~\cite{shsi} carries over to the
quenched case. As a by-product we saw how standard quenched chiral
perturbation theory is recovered.

As discussed in some more detail in Sect.~5.2, for certain
values of the coefficients, the effective potential can
be unbounded from below for large values of the fields.
While the vacuum structure we found satisfies local stability,
this would seem to cast a doubt on our conclusions.  
We believe that this problem is spurious. As we argued,
the effective theory is only valid below the typical hadronic
scale, and should thus not be trusted for large
values of the fields.  In fact, we expect that it is possible to
construct a different effective potential with the same
vacuum structure and the same perturbative expansion,
but which is bounded for large values of the fields
\cite{gh}.

Finally, we remark that the entire analysis was carried
out in euclidean space.  This is the relevant setting:
quenching is only employed in the euclidean version
of lattice QCD.  It is interesting to note that,
while the euclidean ghost
action is invariant under $SO(4)$ rotations, the Minkowski
version would not be invariant under Lorentz transformations,
because in the ghost sector the option of identifying
$\qbar$ with $q^\dagger\gamma^0$ does not exist.  Like
with parity, this would not affect quenched correlation
functions with only physical (and no ghost) quark fields on the
external legs.  However, we doubt that it will
be possible even in principle to continue the quenched
theory to Minkowski space.

\section*{Acknowledgments}

We thank Yigal Shamir for extensive discussions and comments
on the manuscript. We also thank Carleton Detar and Jeff Greensite for 
comments and discussions.
MG and RS thank the Department of Physics and the Institute 
for Nuclear Theory at the University of Washington for hospitality.
MG also thanks the Theory and Applied Physics Divisions (T-8 and X-7) 
at Los Alamos National Lab for hospitality.  SRS thanks the School 
of Physics and Astronomy at the University of Southampton for 
hospitality. All three of us are supported in part by the US 
Department of Energy.

\section*{Appendix}

First, we calculate the small-volume partition
function for the one-flavor quenched theory in the sector with
topological charge zero.  The latter condition implies that
the constant $c_0=0$.  We include this appendix only for
pedagogical reasons; for other cases as well as references,
see for example Refs.~\cite{zirn,jv}.  In a small volume
(often referred to as the ``$\epsilon$-regime"), the dominant
contribution comes from the constant modes.  The 
partition function we wish to calculate is
\ba
\label{PARTF}
Z(m,J)&=&\int_R \omega\;
\Exp\left\{-vV\;\str\left[\pmatrix{m+J&0\cr 0&m}(\Sigma+\Sigma^{-1})
\right]\right\}\ ,\\[5pt]
\Sigma&=&\pmatrix{A&C\cr D&B}\Rightarrow
\Sigma^{-1}=\pmatrix{\frac{1}{A}+\frac{CD}{A^2B}&
-\frac{C}{AB}\cr-\frac{D}{AB}&\frac{1}{B}-\frac{CD}{AB^2}}\ ,\nonumber
\ea
where $V$ is the four-dimensional volume, and the fields are
constant.
$A$ and $B$ represent $c$-number degrees of freedom, while
$C$ and $D$ represent Grassmann degrees of freedom.  The
integration measure $2\pi i\omega=-dA\,dB\,dC\,dD$ is obtained
starting from 
the Haar measure on $GL(1|1)$, and thus, following our
arguments in Sect.~(4.1), we restrict the bosonic part of
the integration region to $R=S_1\times R^+$ \cite{zirn}. In other
words, $A$ lies on the unit circle $S_1$ and $B$ lies on the
positive real axis $R^+$.  We took the masses in the physical
and ghost sectors $m+J$ and $m$ different; the quenched theory
is obtained by setting $J=0$.

A convenient parameterization for the non-linear field $\Sigma$ 
equivalent to that used in the text is given by choosing
\be
\label{SIGMAALT}
A=e^{i\phi'}\ ,\ \ \ B=e^{\phibar'}\ .\\
\ee
The relation between this parameterization and that of 
Eq.~(\ref{PHIPAR}) can be worked out order by order
by an expansion in the fields.  (This expansion is finite, because
of the Grassmann nature of the fields $\chi$, $\chibar$, $C$ and
$D$.)  The advantage of our new parameterization is that the
jacobian which appears when we actually calculate the integral
is simpler.

The jacobian for the transition from $A$, $B$ to $\phi'$,
$\phibar'$ is equal to $iAB=ie^{i\phi'+\phibar'}$.  Expanding
in $C$ and $D$ and defining $r=2vVm$, $s=2vVJ$, we find that
\ba
\label{ZR}
Z(m,J)&=&-\frac{1}{2\pi i}\int_{-\pi}^\pi d\phi'
\int_{-\infty}^\infty d\phibar'\int dC\,dD\,(iAB)\\
&&\Exp\left\{\frac{1}{2}(r+s)(A+1/A)-\frac{1}{2}r(B+1/B)\right\}
\left\{1+\frac{CD}{2AB}\left(\frac{r+s}{A}+\frac{r}{B}\right)
\right\}\nonumber\\
&=&\frac{1}{4\pi}\int_{-\pi}^\pi d\phi'\int_{-\infty}^\infty d\phibar'
\left\{(r+s)\cos{\phi'}+r\cosh{\phibar'}\right\}\nonumber\\
&&\phantom{\frac{1}{4\pi}\int_{-\pi}^\pi d\phi'
\int_{-\infty}^\infty d\phibar'}\times
\Exp\left\{(r+s)\cos{\phi'}-r\cosh{\phibar'}\right\}\nonumber\\
&=&(r+s)I_1(r+s)K_0(r)+rI_0(r+s)K_1(r)\ .\nonumber
\ea
To obtain the quenched result we set $s=0$ and find that
$Z(m,0)=1$, irrespective of the value of $m$.\footnote{Note
that the point $r=s=0$ is singular.}  This would not have been
true if we had chosen $A$ to lie on $R^+$, which corresponds
to choosing $\phi_1=0$ instead of $\phi_2=0$ in Eq.~(\ref{PHIPAR}).

\bigskip

In the case with more than one flavor, there is more than one
way to define the vector-like subgroup $\cal{H}$ of Eq.~(\ref{FULLVECTOR}).
For instance, for the two-flavor case, if we have a mass matrix proportional
to $\tau_3$, it is natural to define the vector-like subgroup
in the ghost sector by requiring that $V$ is $\tau_3$-hermitian:
\be
\label{T3H}
V^\dagger\tau_3 V=\tau_3\ ,
\ee
and likewise, that $V_L=\tau_3 V_R\tau_3$ in the quark sector.
The relevant group integral for the standard quenched two-flavor case,
analogous to the one-flavor case reviewed above,
was done in Ref.~\cite{tv}, and here we discuss only
how things change if one chooses 
vector-like transformations in the ghost sector to
be $\tau_3$-hermitian.

First, let us consider the parameterization of the bosonic ghost
block of $\Sigma$, \ie\ $B$ of Eq.~(\ref{SIGMAALT}) above for the 
two-flavor case.  In Ref.~\cite{tv}, the following parametrization
was chosen (showing only bosonic parameters, \ie\ setting
Grassmann parameters equal to zero):\footnote{We use denote the
parameter $\phi$ of Ref.~\cite{tv} by $\Phi$ here in order to avoid
confusion.}
\be
\label{BPAR}
B=e^{s_1+s_2}\pmatrix{\cosh{\Phi}\;e^{s_1-s_2} & i\sinh{\Phi}\;e^{i\sigma}
\cr i\sinh{\Phi}\;e^{-i\sigma} & \cosh{\Phi}\;e^{-(s_1-s_2)}}\ .
\ee
This can be written as 
\be
\label{BPAROURS}
B=e^{(\phibar_0+\tau_3\phibar_3)/2}\;e^{\tau_1\phibar_1+\tau_2\phibar_2}
\;e^{(\phibar_0+\tau_3\phibar_3)/2}\ ,
\ee
with
\ba
\label{CONNECTION}
s_1+s_2&=&\phibar_0\ ,\\
s_1-s_2&=&\phibar_3\ ,\nonumber\\
\Phi&=&\sqrt{\phibar_1^2+\phibar_2^2}\ ,\nonumber\\
\sin{\sigma}&=&-\frac{\phibar_1}{\Phi}\ ,\ \ \ \ \
\cos{\sigma}=-\frac{\phibar_2}{\Phi}\ .\nonumber
\ea
This parametrizes the coset $GL(2)/U(2)$ for the standard
choice of $\cal{H}$.  For our new definition of $\cal{H}$,
the relevant coset is instead parametrized by
\be
\label{BPARNEW}
B_{\tau_3}=e^{(\phibar_0+\tau_3\phibar_3)/2}
\;e^{i\tau_1\phibar_1+i\tau_2\phibar_2}
\;e^{(\phibar_0+\tau_3\phibar_3)/2}\ ,
\ee
where now the matrix in the middle is unitary instead of hermitian.
Correspondingly, we change the parametrization of the bosonic
valence block, $A$ of Eq.~(\ref{SIGMAALT}), from
\ba
\label{APAR}
A&=&e^{(i\phi_0+i\tau_3\phi_3)/2}\;e^{i\tau_1\phi_1+i\tau_2\phi_2}
\;e^{(i\phi_0+i\tau_3\phi_3)/2}\\
&=&e^{i\psi_1+i\psi_2}\pmatrix{\cos{\theta}\;e^{i\psi_1-i\psi_2} & 
\sin{\theta}\;e^{i\rho}
\cr -\sin{\theta}\;e^{-i\rho} & \cos{\theta}\;e^{-(i\psi_1-i\psi_2)}}\ ,
\nonumber
\ea
in which in terms of the variables of Ref.~\cite{tv}
\ba
\label{ACONNECTION}
\psi_1+\psi_2&=&\phi_0\ ,\\
\psi_1-\psi_2&=&\phi_3\ ,\nonumber\\
\theta&=&\sqrt{\phi_1^2+\phi_2^2}\ ,\nonumber\\
\sin{\rho}&=&\frac{\phi_1}{\theta}\ ,\ \ \ \ \
\cos{\rho}=\frac{\phi_2}{\theta}\ ,\nonumber
\ea
to
\be
\label{APARNEW}
A_{\tau_3}=e^{(i\phi_0+i\tau_3\phi_3)/2}
\;e^{\tau_1\phi_1+\tau_2\phi_2}
\;e^{(i\phi_0+i\tau_3\phi_3)/2}\ .
\ee
In other words, for those generators in which we change the fields
from non-compact to compact in the ghost sector ($\phibar_{1,2}$),
we change the fields from compact to non-compact in the quark
sector (to be compared with the standard choice $\phi_2=0$ in
Eq.~(\ref{PHIPAR})).  

In this parametrization the measures for the matrices $w_{1,2}$
(parametrized by $\phi_{0,3}$ and $\phibar_{0,3}$) and the matrices
$w$ and $\overline{w}$ (parametrized by $\phi_{1,2}$ and $\phibar_{1,2}$)
of Ref.~\cite{tv} factorize.  Following section (3.1.2) of Ref.~\cite{tv},
our change in parametrization corresponds to an interchange of 
the variables $\Phi$ and $\theta$, which leaves the measure 
(\cf\ Eq.~(3.51) of Ref.~\cite{tv}) invariant.  We thus conclude that
if the partition function for the two-flavor case is independent
of the quark mass (as we showed it to be for the one-flavor case
above), it is also independent of the quark mass for our new
parametrization, if we follow the prescription given
above.

\end{document}